# Highly Conductive Tungsten Suboxide Nanotubes.

Cécile Huez,[1] Maxime Berthe,[1] Florence Volatron,[2] Jean-Michel Guigner,[3] Dalil Brouri,[4] Lise-Marie Chamoreau,[2] Benoît Baptiste,[3]
Anna Proust[2,*] & Dominique Vuillaume.[1,*]

1) Institute for Electronics Microelectronics and Nanotechnology (IEMN), CNRS, University of Lille, Av. Poincaré, Villeneuve d'Ascq, France
2) Institut Parisien de Chimie Moléculaire (IPCM), CNRS, Sorbonne Université, 4 Place Jussieu, F-75005 Paris, France
3) Institut de Minéralogie, de Physique des Matériaux et de Cosmologie (IMPMC), CNRS, Sorbonne Université, Muséum National d'Histoire Naturelle, 4 place Jussieu, F-75005 Paris, France
4) Laboratoire de Réactivité des Surfaces (LRS), CNRS, Sorbonne Université, 4 place Jussieu, F-75005 Paris, France

* Corresponding authors: dominique.vuillaume@iemn.fr ; anna.proust@sorbonne-universite.fr

## ABSTRACT.

We demonstrate a high electron conductivity (>$10^2$ S/cm and up to $10^3$ S/cm) of tungsten suboxide $W_{18}O_{52.4-52.9}$ (or equivalently $WO_{2.91-2.94}$) nanotubes (2-3 nm in diameter, ~ µm long). The conductivity is measured in the temperature range 120 to 300K by a four-probe scanning tunneling microscope in ultra-high vacuum. The nanotubes are synthesized by a low-temperature and low-cost solvothermal method. They self-assemble in bundles of hundreds of nanotubes forming nanowires (~ µm long, few tens nm wide). We observe a large anisotropy of the conductivity with a ratio (longitudinal conductivity/perpendicular conductivity) of ~ $10^5$. A large fraction of them (~ 65-95%) shows a metallic-like, thermal activation less, electron transport behavior. Few of them, with a lower conductivity from 10 to $10^2$ S/cm, display a variable range hopping behavior. In this latter case, a hopping barrier energy of ~ 0.24 eV is inferred in agreement with the calculated energy level of the oxygen vacancy below the conduction band. This result is in agreement with a relative average concentration of oxygen vacancies of ~ 3%, for which a semiconductor-to-metal transition was theoretically predicted. These tungsten suboxide nanostructures are prone to a wide range of applications in nanoelectronics.

# I. INTRODUCTION

Nanotubes (NTs) and nanowires (NWs) are nanostructures with plenty of properties for multiple potential applications. For example, metallic NTs and NWs are of uppermost importance for semi-transparent electrodes, connecting vias in microelectronic chips, chemical sensors.[1] Focusing on NTs and NWs with a high electrical conductivity, multi-wall carbon nanotubes (MWCNT) are metallic with a conductivity at room temperature (RT) in the range $10^3$-$10^5$ S/cm.[2] The RT conductivity of organic NWs based on conducting polymers (*e.g.*, poly(3,4-ethylene dioxythiophene) (PEDOT), polyaniline (PANI), polypyrrole (PPy)) were reported between $10^{-1}$ and $10^3$ S/cm depending on doping level, polymer structure, presence of disorder and defects, diameter.[3, 4] A conductivity between ~2 and ~10 S/cm was measured at RT for metal coordination polymers nanoribbons (*e.g.*, $[Pt_2I(S_2CCH_3)_4]_n$).[5] Metal oxide-based NWs obtained by different fabrication processes (CVD growth, pyrolysis-reduction process, thermal evaporation,…) were also studied. Tungsten suboxide $W_{18}O_{49}$ (or equivalently $WO_{2.72}$) NWs were synthesized, and values of electrical conductivity between ~2 and ~15 S/cm were reported at RT with a semiconducting behavior.[6-11] However, *ab initio* calculations predicted that all the substoichiometric phases ($WO_x$, 2.625<x<2.92, e.g., $W_{32}O_{84}$, $W_{18}O_{49}$, $W_{17}O_{47}$, $W_5O_{14}$, etc...), known as Magnéli phases of tungsten oxides, should have a metal-like behavior.[12] The advantages of tungsten suboxide nanostructures, owing to the variety of their stoichiometric phases and crystal structures, rely on their wide range of potential applications in photochromic materials for smart windows, conducting transparent electrodes, near-infrared shields, optoelectronics, gas sensors, data storage devices, supercapacitors, iontronic devices, and even more.[13-18]

Here, we synthesized by solvothermal methods $WO_x$ NWs (here 2.91<x<2.94, *vide infra*) constituted by bundles of hundreds of NTs. Solvothermal synthesis is a low temperature method with specific pressure conditions that change some properties of the solvents and precursors. This technique generally induces the formation of species of unprecedented nature and morphology.[19, 20] We demonstrated using four-probe scanning tunneling microscope (4P-STM) in ultra-high vacuum (UHV), an unprecedented electrical conductivity (up to $10^3$ S/cm at room temperature) along them. These results outperform previously reported results (~ 2 to 15 S/cm at RT) for tungsten suboxide nanostructures.[6-11]

A metallic-like behavior is demonstrated for a large fraction (65-95%) of these bundles of NTs by temperature-dependent (120-300K) measurements. We observed an activation-less electronic transport properties and an absence of a gate-voltage modulation of the transport along them. A high anisotropy is measured between the longitudinal, $\sigma_L$, and the perpendicular, $\sigma_P$, conductivities (ratio $\sigma_L/\sigma_P \approx 10^5$). Finally, we also observed that a small number of bundles, with lower conductivity between 10 and $10^2$ S/cm, exhibit a temperature-dependent behavior. This feature was accounted for by a variable range hopping model, with a hopping barrier energy of ~ 0.24 eV, in agreement with the calculated energy level for an oxygen vacancy in tungsten suboxides.[21] A semiconducting-to-metal transition was theoretically predicted in tungsten suboxides when the concentration of oxygen vacancies increases in the range 2 to 4%.[21] We assume that the observed coexistence of metallic-like and semiconducting-like tungsten suboxides is likely due to the dispersion of the concentration of these oxygen vacancies (here around a mean value of 3%) in our bundles of NTs.

## II. RESULTS AND DISCUSSION

### A. Structural and physico-chemical characterizations.

The WO$_x$ NWs were produced by a solvothermal method following published procedures for the synthesis of W$_{18}$O$_{49}$ (or equivalently WO$_{2.72}$),[13, 22] see details in the Supplementary Material. The synthesized deep blue powder was dispersed by drop casting on a substrate. The tapping-mode atomic force microscopy (TM-AFM) and the transmission electron microscopy (TEM) images show that the deposited NWs have an average length of 0.8-1.5 µm and an average diameter of few tens of nm (Figs. 1a-d), details of the TM-AFM and TEM methods in the Supplementary Material. All the NWs have a spindle shape with a typical height/width of ~60/110 nm on the center and ~25/50 nm on the ends (Fig. 1e) in agreement with Ref. 13.

This spindle shape is observed whatever the nature of the underlying substrate: amorphous carbon (TEM grid), SiO$_2$ and Au for TM-AFM (Figs. 1a-c, and Fig. S1 in the Supplementary Material). The bright field TEM micrographs (Figs. 2a-e, and additional Fig. S2 in the Supplementary Material) reveal that the NWs are composed of an assembly of nanotubes (NTs), which was not previously reported for these tungsten suboxide nanostructures.[13, 22-24] Indeed, the amplitude contrast of these images (dark at the edges and light inside) is consistent with the "hollow"-like structure of nanotubes, as also reported for nanotubes of several other materials.[25, 26] For these single nanostructures (marked by the yellow arrows in Figs. 2b, d and e), we observed two parallel black lines starting and ending at the same position. The distance between the two parallel black lines is constant (about 2-2.5 nm) all along the single nanostructure. This feature is in favor of the walls of a single NT rather than to two individual nanofibers placed side-by-side. Thus, we suggest that the NWs be made of bundles of NTs. From a careful analysis of many images, these NTs have an external diameter $d_{nt}$ 2.7 ± 0.7 nm (distance between the two walls, dark gray) and an internal diameter $\delta_{nt}$ (light gray area between two walls) of 1 ± 0.35 nm (dataset in Table S1 in the Supplementary Material). High resolution observations (HRTEM) were carried out and the HRTEM micrographs (Fig. 2f) point out the crystalline nature of the NTs. The micrographs were recorded in optimal conditions (orientation of the sample and defocus of the objective lens) to observe the atomic planes (phase contrast). The atomic planes (Fig. 2f) are perpendicular to the NT long axis and they have a lateral extension corresponding to the width of the NT, around 3 nm. The fringes and FFT analysis (inset in Fig. 2f) of the high-resolution TEM images (see also Fig. S2d and S2f in the Supplementary Material) indicate a crystalline structure with an interplanar spacing b value between 0.36 and 0.39 nm along the growth direction (Table S2 in the Supplementary Material) in good agreement with other results and consistent with the (010) plane of the monoclinic $W_{18}O_{49}$ ($WO_{2.72}$) and other tungsten suboxides ($WO_{2.92}$, $WO_{2.90}$).[22, 27]

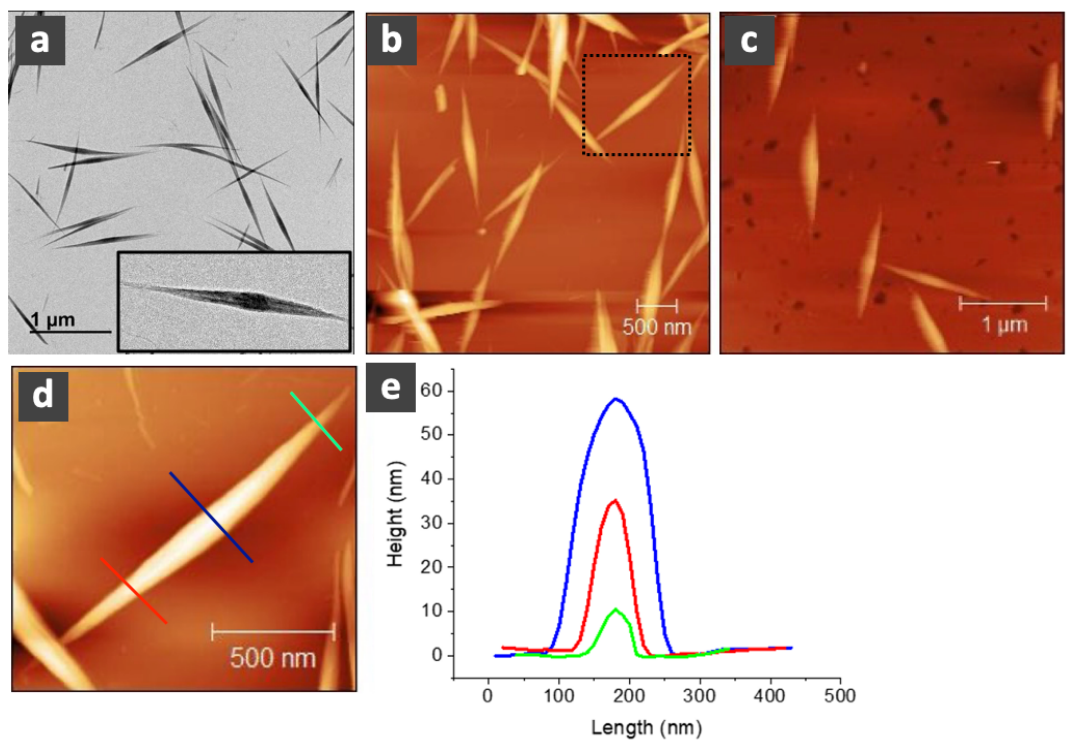

*Figure 1.* (a) TEM image of NWs deposited on carbon, inset : zoom on a single NW. Topographic TM-AFM of NWs deposited on (b) $SiO_2$ and (c) Au (dark spots

*are tiny pinholes in the Au surface). (d) Zoom on a NW on SiO$_2$ (dashed lines in (b)) and (e) profiles at 3 locations along the NW.*

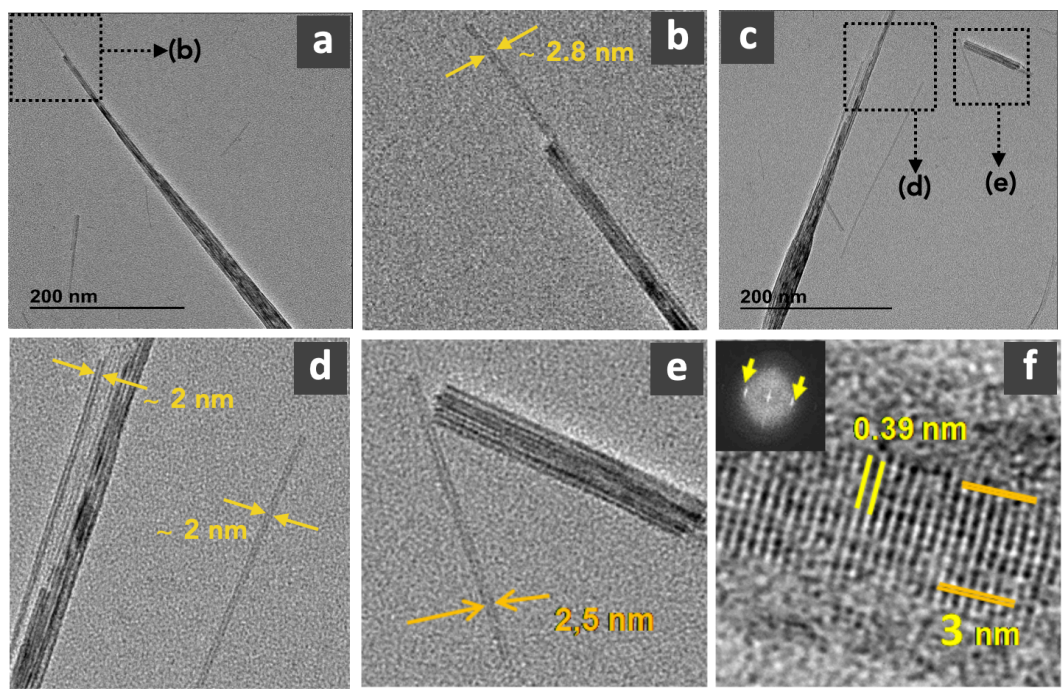

*Figure 2. TEM images showing the NTs structure of the NWs. (a) and (c) Images of the ends of two NWs. (b, d and e) Zooms of the areas indicated by dashed lines in (a) and (c), respectively, showing the assembly of NTs and the estimate of their diameters. (f) HR-TEM of a single NT with an interplanar spacing of 0.39 nm along the growth direction. The inset shows the corresponding FFT, where the two spots enable the calculation of the inter-reticular distance 3.9 Å. See protocol details in section S3 in the Supplementary Material.*

The formation of the tungsten suboxide phase is confirmed by the X-ray diffraction (XRD) pattern (Fig. 3a), which shows the expected narrow (010) peak in good agreement with previous reports on the structural characterization of anisotropic nano-objects of tungsten suboxides.[11, 13, 22] The XRD pattern (black open circles, Fig. 3a) is well fitted with the $W_{18}O_{49}$ ICSD-15254 crystal structure data.[28] We note that due to the fact that we have obtained highly anisotropic nanostructures randomly oriented, as evidenced by TEM and TM-AFM images (Fig. 1), only a very few peaks are sharp compared to XRD data for bulky crystal suboxide materials reported in the literature.[7, 9, 11, 29, 30] Le Bail refinement[31] of the diffractogram leads to a *b* value of 0.3796(2) nm (Table S3 in the Supplementary Material). However, it is also possible to model the XRD experimental data with other tungsten suboxides of various compositions, even $WO_3$ (see Fig. S3 in the Supplementary Material). Thus, we cannot determine the precise stoichiometry of the compound from XRD experiments, as already stated in previous work on tungsten suboxide nano-objects.[32] The Raman and XPS analysis (Figs. 3b-d) confirm the chemical composition of the NWs. The Raman spectrum (Fig. 3b) displays the characteristic peaks of tungsten suboxide, at 73 and 122 cm$^{-1}$ (W-O-W bending modes), 218 and 345 cm$^{-1}$ (O-W-O bending modes), 674 and 803 cm$^{-1}$ (W-O stretching modes).[11, 13, 22, 33] The XPS spectrum can be decomposed into two W $4f_{5/2}$-$4f_{7/2}$ doublets, one for the $W^{6+}$ oxidation state (38.3 and 36.1 eV) and the other one for the $W^{5+}$ oxidation state (36.8 and 34.6 eV). A contribution from the W $5p_{3/2}$ peak is also observed as previously reported (section S5 and Table S4 in the Supplementary Material).[29] These oxidation states are in agreement with previously reported results for tungsten suboxides (except a very weak $W^{4+}$ not detected here).[11, 13, 22] The O 1s core level (Fig. 3d) shows a broad main peak (at 530.6 eV) associated to W-O bonds with additional weak peaks at higher binding energies (531.6 and 533.6 eV) likely associated to residual contamination (water, C-O bonds in contaminants).[22] We carried out the same XPS analysis on several series of samples (freshly prepared samples and after more than 19 months) and we obtained the same main features with only small variations of the relative amplitude of the $W^{6+}$ and $W^{5+}$ peaks from sample-to-sample, which likely reflect small dispersion of the stoichiometry (see section 5 in the Supplementary

Material). From the integration of the W 4f doublets, we estimated a stoichiometry of the nanostructures between $WO_{2.91}$ and $WO_{2.94}$ for the various measured samples (Fig. 3c and Fig. S4), the ideal $W_{18}O_{49}$ being equivalent to $WO_{2.72}$. This indicates an oxygen enrichment of these nanostructures as also reported in Ref. 13. We conclude that the synthesis of these tungsten suboxide is reasonably reproducible and the structure of the nanostructures shows a long-term stability under ambient conditions.

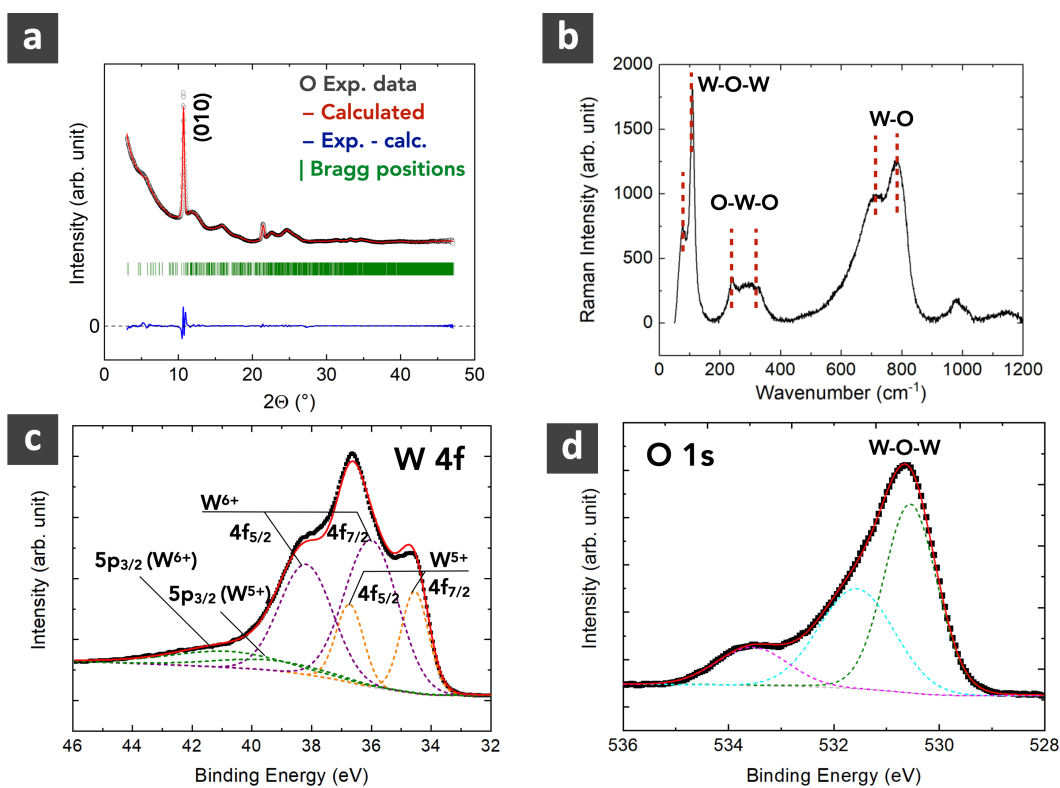

*Figure 3.* (a) XRD pattern (Mo-Kα radiation): the measured pattern (black open circles), the calculated pattern (red line), and the difference between the two (blue line). The green bars show the expected positions of Bragg peaks related to $W_{18}O_{49}$ structure (ICSD-15254 crystal structure data, see more details in the Supplementary Material, section S4). (b) The Raman spectrum showing the *tungsten oxide contributions. (c, d) XPS spectra (experimental data: black squares; fit: red line) of the $W_{4f}$, $W_{5p}$ and $O_{1s}$ peaks and their deconvolutions (dashed lines) (see details section S5 in the Supplementary Material).*

## B. Longitudinal conductivity at room temperature.

In the following, NW is used to refer to the properties of the bundle of NTs, and NT used to refer to the properties of a single NT inside the bundle. The electrical conductivity along an individual NW deposited on a Si n$^+$/SiO$_2$ substrate (highly doped Si with a resistivity of 1-3x10$^{-3}$ Ω.cm, 200 nm thick thermally grown SiO$_2$) was measured in UHV using a 4-probe STM (4P-STM, protocol details in section S6 in the Supplementary Material, the Fig. 4a is a scheme of the measurement). Fig. 4b shows a dataset of voltage-current (V-I) curves measured at room temperature (RT). We have measured 52 V-I curves on 29 NWs (with a zero bias on the underlying Si). The longitudinal NW conductance, $\sigma_{Lnw}$, was determined for each individual NW by $\sigma_{Lnw}=G_L(L/S_{nw})$ from the measured conductance $G_L$ of the NW (inverse of the slope of the V-I curves). L is the inner-probe distance. $S_{nw}$ is the cross-section area of the NW that was systematically estimated for each NW by measuring the NW diameter (at its thickest part in the middle) with the scanning electron microscope (SEM) installed inside the 4P-STM (detailed datasets in Table S5 in the Supplementary Material). This approximation of using a cylinder shape, instead of the real spindle shape, to calculate the conductivity is discussed in the Supplementary Material (section S6). In brief, the induced error was estimated well below the data dispersion, thus without a significant impact on the conclusions. Note also that the V-I measurements were acquired with the e-beam of the SEM turned off and we carefully checked that the e-beam irradiation during the positioning of the STM probes on the NWs has a weak effect. A slight increase by ~ 1.5 was observed on the electrical behavior of the NWs (see section 7 in the Supplementary Material). These conductivity values are "effective" conductivities neglecting voids inside the NTs and between adjacent NTs in the bundle. The longitudinal conductivity of a single NT inside the NW is estimated by $\sigma_{Lnt}=G_L(L/S_{nt})/N$. $S_{nt}$ is the cross-section surface of a NT given $S_{nt}= \pi(d_{nt}^2-\delta_{nt}^2)/4$ where $d_{nt}$ is the external diameter, $\delta_{nt}$ the internal diameter, and N the number of single NT in the bundle (detailed datasets in Table S5 in the Supplementary Material). We used the mean values $d_{nt}$=2.7 nm (± 0.7 nm) and $\delta_{nt}$=1 nm (± 0.3 nm) as determined from TEM (*vide supra,* Table S1 in the Supplementary Material). N is estimated for each bundle from the NW diameter ($d_{nw}$) and the NT external diameter $d_{nt}$ by N≈ $(d_{nw}/d_{nt})^2$ (assuming a cylindrical NW). We get values of $\sigma_{Lnt}$ of the same order of magnitude, but 15-20% larger than above for the NW (on average) - Table S5 in the Supplementary Material. Figures. 4c and 4d show the histograms of the conductivity values $\sigma_{Lnw}$ and $\sigma_{Lnt}$, respectively. The values are log-normal distributed with a mean value $\overline{\sigma_{Lnt}}$ ≈ 510 S/cm and $\overline{\sigma_{Lnw}}$ ≈ 440 S/cm. At RT, we observed a distribution of the single NT conductivity values $\sigma_{Lnt}$ between ≈ $10^2$ and $10^3$ S/cm (Fig. 4d) with a majority of values (>95%) above $10^2$ S/cm.

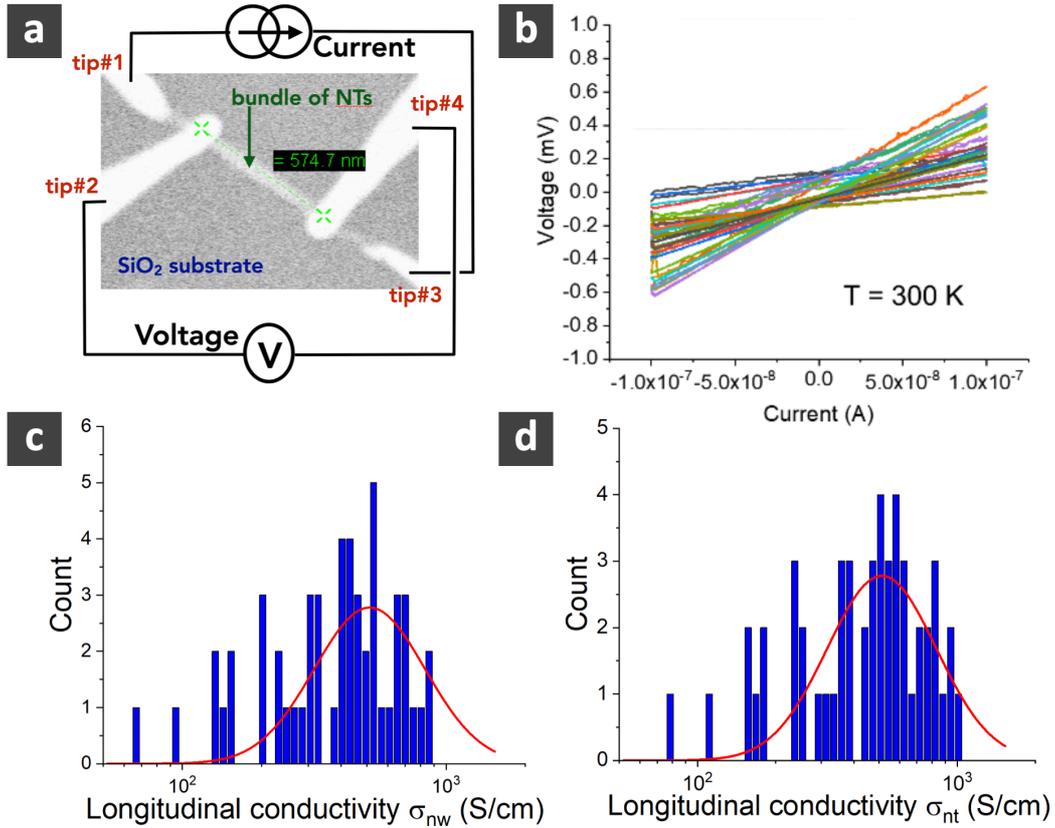

*Figure 4.* *(a) Scheme of the 4-probe measurement of the longitudinal conductivity. The bundle of NTs (SEM image) is connected by the 4 tips (details in section S6 in the Supplementary Material). The current is applied on the external probes (tip#1 and tip#3) and the resulting voltage is measured between the inner probes (tips#2 and tip#4 separated by ~575 nm in this case). This 4-probe technique eliminates the effect of the contact resistance from the measured sample conductance, unlike the usual 2-probe measurements. (b) V-I dataset at RT (52 V-I traces measured on 29 NWs). We note a voltage offset at zero current due to an internal offset of the voltage amplifier used for the measurement. Since only the slope of the V-I traces is important to calculate the conductivity, this offset was not corrected. The histograms of (c) $\sigma_{Lnw}$ and (d) $\sigma_{Lnt}$ calculated from the dataset in Fig. 4b (and data details in Table S5 in the Supplementary Material). The red lines are fits by a log-normal distribution with mean values of*

$\overline{\sigma_{Lnw}} \approx 440$ S/cm  (log-mean=2.64, log-standard deviation=0.22) and $\overline{\sigma_{Lnt}} \approx 510$ S/cm (log-mean=2.71, log-standard deviation=0.22).

## C. Temperature-dependent conductivity.

The same measurements were performed down to 120K. We clearly observed a broadening of the $\sigma_{Lnt}$ values, between 10 and $10^3$ S/cm, with a majority of values (from 65% at 120K to 95% at RT above the red-dash line in Fig. 5a, i.e., $\sigma_{Lnt} > 10^2$ S/cm) showing an activation-less temperature behavior (V-I datasets for each temperature are given in Fig. S6 and Table S6 in the Supplementary Material). However, a fraction of the NTs, with conductivity $< 10^2$ S/cm between the red-dash line and the black dash line in Fig. 5a, displayed a temperature-dependent behavior. For these non-metal-like NTs, several transport mechanisms can be considered: a classical temperature-activated transport mechanism (Arrhenius), a polaron hopping transport as suggested for $W_{18}O_{49}$ nanowires synthesized by a low temperature (600°C) furnace process.[7] If we consider the limit of the lowest conductivity values in the distribution for each temperature (black dash line in Fig. 5a), these data were not well fitted by these models (section 8 in the Supplementary Material), given nonsignificant low activation energies (tens of meV , *i.e.*, of the order of kT). These temperature-dependent mechanisms were discarded. These data better follow the Mott VRH (variable range hopping) law, $\sigma = \sigma_0 \exp(-(T_M/T)^{1/4})$ (Fig. 5b), where $T_M$ is the Mott temperature (*vide infra*). This model characterizes electron transport in disordered semiconductors and amorphous solids.[34, 35] We conclude that a fraction of the NTs, with the lowest conductivity, have a more disordered atomic structure or a slightly different concentration of oxygen vacancies (*vide infra*). We note that the mean longitudinal conductivity is not significantly dependent on the temperature (Fig. S8-c in the Supplementary Material), fluctuating between $\approx 200$ (at 120 K) and 450 S/cm (at RT). This feature is consistent with the largest fraction of metallic-like NTs.

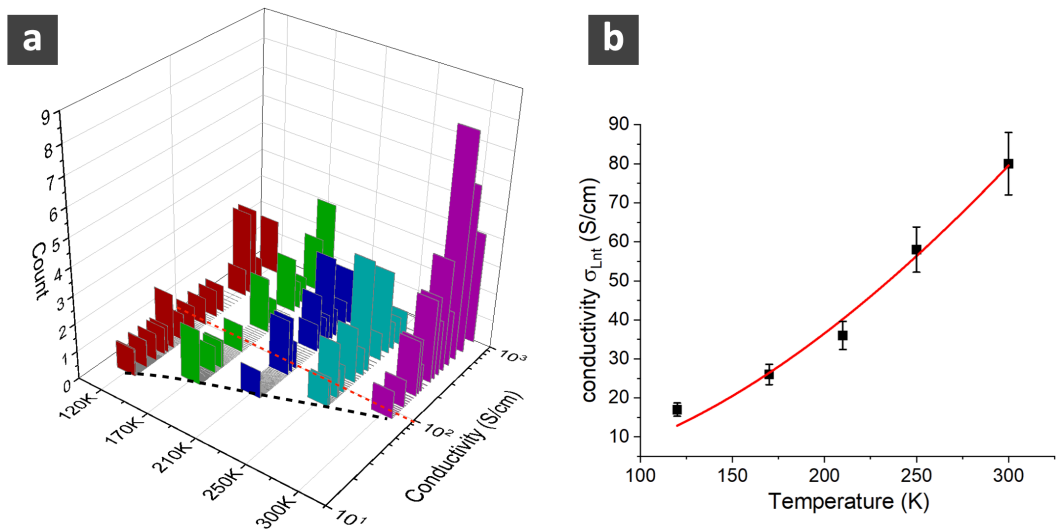

*Figure 5*. (a) *Histograms of the temperature-dependent measurements of the NT conductivity σ$_{Lnt}$. (b) Plot of the lowest conductivity (squares are the values taken along the black dashed line in Fig. 5a) versus temperature and fit (red line) with a VRH model (σ$_0$=1.1x10$^5$ S/cm, T$_M$=8.5x10$^5$ K with r$^2$=0.98).*

Considering the high σ$_{Lnt}$ values (> 10$^2$, up to 10$^3$ S/cm) and the thermal activation-less behavior, we conclude that a large fraction of these W$_{18}$O$_{52.4-52.9}$ NTs have a metallic-like character. This behavior is confirmed by the absence of a field effect on the electron transport in the NW when applying a bias voltage on the underlying highly doped Si (gate voltage in a field-effect transistor configuration, section S9 in the Supplementary Material). These findings outperform previously reported performances of most of the tungsten suboxide W$_{18}$O$_{49}$ NWs (other preparation methods, not solvothermal but at high temperatures), *i.e.,* with a semiconducting behavior and a conductivity lower than 15 S/cm.[6-11] We note, however, that a value of conductivity up to 4x10$^4$ S/cm has been reported but for more macroscopic filled NWs (240 nm and 310 nm in diameter). However, in this latter case, we cannot exclude some metal contamination (e.g. Ni) resulting from the synthesis protocol.[36] Thus, a comparison with our data can be discarded. Our results are on par with those of NWs of the archetype conducting polymer PEDOT:PSS and other conducting polymers (PANI, PPY).[4]

It is known from *ab initio* calculations that the precise stoichiometry of the tungsten suboxides is important to dictate the metallic or semiconducting behaviors[12, 37] and that a semiconducting-to-metal transition exists when increasing the concentration of oxygen vacancies ($V_O$) in this material in the range 2-4% ($V_O$ concentration calculated by reference to $WO_3$).[21] In our case, the stoichiometry determined by XPS (*vide supra*) corresponds to a relative concentration of $V_O \approx$ 2-3% (i.e., the percentage of missing O with respect to the $WO_3$ stoichiometry). The experimentally observed coexistence of metallic-like and semiconducting-like tungsten suboxide NTs can be rationalized assuming the dispersion of the concentration of $V_O$ in the synthesized NTs.

For the fraction of semiconducting NTs following a Mott VRH electron transport model, the Mott temperature, $T_M \approx 8.5 \times 10^5$ K (from the fit in Fig. 5b) is related to several parameters of the hopping sites by $T_M = (24/\pi)\alpha^3/kN(\varepsilon_F)$,[34, 38] with k the Boltzmann constant, $N(\varepsilon_F)$ the density of states at the Fermi energy, and α a parameter characterized by the integration of all possible tunneling paths between two hopping sites, *i.e.*, the α parameter reflects the "potential landscape" surrounding the hopping sites. A value $N(\varepsilon_F) \approx 5 \times 10^{21}$ $eV^{-1} cm^{-3}$ has been estimated from *ab initio* calculations,[12, 21] we deduced $\alpha \approx 3.6$ $nm^{-1}$ from the estimated $T_M$. In a very simplified picture, considering tunneling between two adjacent hopping sites, α can be approximated by the tunneling decay factor $\alpha=2(2m^*\Delta)^{1/2}/\hbar$, with m* the electron effective mass, ℏ the reduced Planck constant and Δ the tunneling energy barrier.[39, 40] If we assume an effective mass $m^*=0.5m_0$ ($m_0$ the electron rest mass in the vacuum), we deduced a hopping tunnel barrier of Δ ≈ 0.24 eV. Albeit, this estimate is clearly simplified, we note that this energy value is consistent with the calculated energy level of $V_0$ in tungsten suboxides at around 0.1 - 0.25 eV below the conduction band.[21]

### D. Perpendicular conductivity.

The conductivity perpendicular to the long axis of the NWs was measured by Conducting-AFM (C-AFM) at RT (details in section S2 in the Supplementary Material). The figure 6a shows the current-voltage (I-V) curves acquired on several NWs deposited on ultra-flat template-stripped $^{TS}$Au substrates (rms roughness ~ 0.4 nm)[41-43] - datasets in section 10 in the Supplementary Material. From the slope of the I-V curves around 0 V (i.e., ±50 mV), we estimated the zero-bias conductance $G_p$ and we calculated the perpendicular conductivity $\sigma_P = G_p(S_c/d_{nw})$. We estimate a NW diameter (height) $d_{nw}$ ~ 60 nm from the topographic AFM image shown in the inset Fig. 6a and a C-AFM tip surface contact $S_c$ ~ 39 nm$^2$ using a mechanical Hertz model[44] and a measured Young modulus of the tungsten oxide NWs,[45] see Section S11 in the Supplementary Material. Fig. 6b shows that the values of $\sigma_P$ are largely dispersed, and we deduced a mean perpendicular conductivity of $\overline{\sigma_P}$ ≈ 6x10$^{-4}$ S/cm. We note that these values measured in ambient air are stable over several hours (time of the experiments) and not sensitive to degradation upon air exposure[24, 46] as observed for the longitudinal transport. Taking into account the correction factor with respect to the 4-probe STM experimental conditions (e-beam exposure, *vide supra* and section 7 in the Supplementary Material), the mean conductivity anisotropy $\overline{(\sigma_{Ln}/1.5)/\sigma_P}$ for the NWs is estimated to be ≈10$^5$. It is worth noting that this result outperforms the anisotropy of conductivity for conducting polymer (PPy, PEDOT, PANI,...) and CNT nanosheets and nanowires reported between ≈ 50 and 10$^4$.[47-51] This result is understood because of the very different transport mechanisms in the two cases (Fig. S10 in the Supplementary Material). In the longitudinal direction, the electron transport (for metal-like NTs) is likely due to drift-diffusion along the individual NT, with only a few hopping between adjacent NTs in the bundle. Thus, the topology of ET in the bundle can be simply viewed as conducting channels in parallel. On the opposite, the electron transport across the NWs is mainly due to tunnel hopping between neighboring NTs and the topology of the ET pathways is more complex depending on how exactly a NT interacts with its neighboring NTs and how many they are around (like a 2D percolation network). In such a complex ET network, the overall conductivity is somehow limited by the less efficient ET channel. In this latter case, this ET mechanism also explains the larger dispersion of the perpendicular conductivity (Fig. 6b) compared to the dispersion of the longitudinal conductivity (Fig. 4). This larger dispersion is likely due to a greater sensitivity of the overall electron transport to the precise organization of between the adjacent NTs and consequently to large variations of the tunnel hopping probabilities, impacting the measured perpendicular conductance. Compared to the longitudinal transport for which the electrons are transported along individual NTs in parallel without the need for a strong interaction between them, the measured dispersion mainly reflects the intrinsic dispersion of the NTs related to the fluctuations of the chemical stoichiometry - see section C. We also note that a part of this larger dispersion of the perpendicular conductivity can come from the variations of the two contact resistances (at the tip/bundle and bundle/Au substrate, *e.g.* due to fluctuations of the C-AFM tip loading force), which are not taken into account in this two-probe configuration (while the effects of the contact resistances are eliminated by the 4-probe configuration used for the measurement of the longitudinal conductivity).

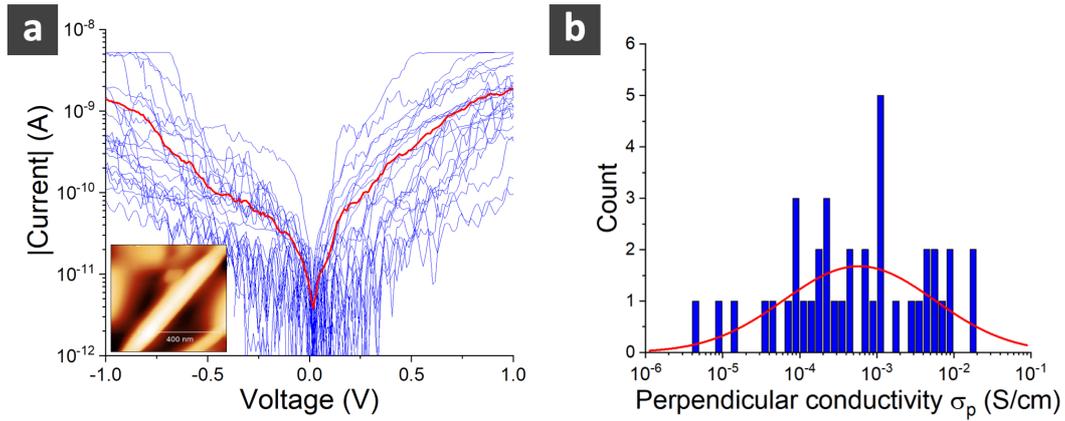

*Figure 6*. *(a) Dataset of the perpendicular current (absolute value, log scale) versus voltage (C-AFM at RT in ambient condition, 41 I-Vs) of NWs deposited on $^{TS}$Au. The bold red line is the mean current curve. (b) Histogram of the perpendicular conductivity, $\sigma_P$ and fit with a log-normal distribution with a mean value $\overline{\sigma_P}$ = 6x10$^{-4}$ S/cm (log-mean=-3.23, log-standard deviation=0.95).*

## III. CONCLUSION

In conclusion:

- Tungsten suboxide nanotubes with a diameter of 2-3 nm and a length of few µm were synthesized by a low temperature, low cost, solvothermal method.

- These nanotubes self-assemble to form µm long nanowires (bundles of nanotubes) with diameters between 20 and 100 nm. We note that only solid nanowires (not bundle of NTs) were reported previously using similar synthesis routes.[13, 22, 24] Whatever the origins of this feature could be, *e.g.*, subtle changes of specific details of the synthesis, the present results clearly call for more studies devoted to the growth mechanisms and structural characterizations of these suboxide nanotubes and how they self-assemble in bundles, out of the scope of this work.

- A large fraction (~65%-95%) of these nanotubes have unprecedented high longitudinal electrical conductivity ($10^2$-$10^3$ S/cm) with a metallic-like behavior (thermal activation-less electron transport) as revealed by temperature-dependent 4-probe STM measurements in UHV. This high, metal-like, electrical conductivity is consistent with theoretical prediction for Magnéli phases of this material.
- A small fraction of them, with conductivity in the range 10 to $10^2$ S/cm at low temperatures, follow a variable range hopping behavior, with a hopping energy barrier of ca. 0.24 eV.
- This feature is understood by considering that the stoichiometry of these nanotubes corresponds to a concentration of oxygen vacancies of ~2-3%, a value for which a semiconducting-to-metal transition has been theoretically predicted.[21]
- These tungsten suboxide nanowires have a high anisotropic conductivity behavior with a ratio ≈ $10^5$ between their longitudinal and transversal (though the bundle of nanotubes) conductivities.

These highly conductive tungsten suboxide nanostructures are prone for applications in electrochromic devices and smart windows,[52, 53] transparent conducting electrodes,[54] gas sensors,[55, 56] and field-emission electron source devices.[30, 57-59] More exploratory, we can envision applications in neuromorphic devices and brain-inspired computing systems based on two-dimensional networks of dense and interconnected nanowires, eventually decorated with molecules, as recently simulated[60, 61] and explored with sulfured silver nanowires,[62, 63] carbon NT,[64, 65] or crossbar networks of memristive devices based on thin films of $WO_3$.[66-69] The high anisotropy of conduction is also a key advantage that gives a device dual properties along the two orthogonal directions. For example,[70] if the relative variation of conductivity upon gas exposure displays an anisotropic responsiveness and a different sensitivity to a specific gas in the two orthogonal directions, this can be used to develop a highly selective gas sensor. Similarly, an anisotropic current response upon a mechanical strain constraint of the bundle of NTs can be applied to a directional strain sensor (see a discussion of more device applications endowed by anisotropy of properties in a recent review paper, Ref. 70).

## SUPPLEMENTARY MATERIAL

See the supplementary materials for details on synthesis, TEM, HR-TEM, XPS, C-AFM, 4P-STM characterizations, complete set of conductivity data by 4P-STM and C-AFM, and control experiments.

## ACKNOWLEDGEMENTS.

We acknowledge funding by the CNRS (France), project neuroPOM, under a grant of the 80PRIME program. We acknowledge Xavier Wallart (IEMN-CNRS) and Michel Daher-Mansour (IEMN-CNRS) for their help with the XPS and Raman spectroscopy measurements, respectively.

## AUTHOR DECLARATIONS

**Conflict of Interest**

The authors have no conflicts to disclose.

**Author Contributions**

**Cécile Huez**: Investigation (equal); Data curation(equal); Formal analysis (equal); Writing-review & editing (equal). **Maxime Berthe**: Investigation (equal); Writing-review & editing (supporting). **Florence Volatron**: Investigation (supporting); Formal analysis (supporting); Writing-review & editing (supporting). **Jean-Michel**

**Guigner**: Investigation (equal); Formal analysis (equal); Writing-review & editing (supporting). **Dalil Brouri**: Investigation (equal); Formal analysis (equal); Writing-review & editing (supporting). **Lise-Marie Chamoreau**: Investigation (equal); Formal analysis (equal); Writing-review & editing (supporting). **Benoît Baptiste**: Investigation (equal); Formal analysis (equal); Writing-review & editing (supporting). **Anna Proust**: Conceptualization (equal); Funding acquisition (equal); Supervision (lead); Project administration (equal); Writing-original draft (equal); Writing-review & editing (equal). **Dominique Vuillaume**: Conceptualization (equal); Formal analysis (lead); Funding acquisition (equal); Supervision (lead); Project administration (equal); Writing-original draft (lead); Writing-review & editing (lead).

## DATA AVAILABILITY

The data that support the findings of this study are available from the corresponding author upon reasonable request.

## REFERENCES.

# Highly Conductive Tungsten Suboxide Nanotubes.

Cécile Huez,[1] Maxime Berthe,[1] Florence Volatron,[2] Jean-Michel Guigner,[3] Dalil Brouri,[4] Lise-Marie Chamoreau,[2] Benoît Baptiste,[3]
Anna Proust[2,*] & Dominique Vuillaume.[1,*]

1) Institute for Electronics Microelectronics and Nanotechnology (IEMN), CNRS, University of Lille, Av. Poincaré, Villeneuve d'Ascq, France

2) Institut Parisien de Chimie Moléculaire (IPCM), CNRS, Sorbonne Université, 4 Place Jussieu, F-75005 Paris, France

3) Institut de Minéralogie, de Physique des Matériaux et de Cosmologie (IMPMC), CNRS, Sorbonne Université, Muséum National d'Histoire Naturelle, 4 place Jussieu, F-75005 Paris, France

4) Laboratoire de Réactivité des Surfaces (LRS), CNRS, Sorbonne Université, 4 place Jussieu, F-75005 Paris, France

* Corresponding authors: dominique.vuillaume@iemn.fr ; anna.proust@sorbonne-universite.fr

# Supplementary Material

## S1. Synthesis.

276.5 mg of $WCl_6$ was weighed into a Teflon box in a glove box, then 60 mL of propanol was added under argon flow funnel (concentration 11.6 mM). The solution was mixed and then sealed in the autoclave (Parr Model 4748, 125 mL). The solvothermal synthesis started by letting the solution 10 minutes at 25°C, then heated during 1h to reach 180°C and maintained 24h at this temperature. Finally, the solution was slowly cooled at 25°C during 18h. The product was decanted with a plastic pipette in a centrifugation tube (previously purged). We obtained a blue powder. We centrifuged first at 5000 rpm during 5 min and then we washed twice with $H_2O$ (6000 rpm during 6 min and 9000 rpm during 10 min). We finished by a third wash with ethanol at 9000 rpm during 10 min. Both solvents were also previously purged. We dried under vacuum at 50-60°C during few hours and let under vacuum without heating during the whole night. We obtained 110 mg of a dark blue powder.

## S2. AFM and C-AFM characterizations.

All the topographic and conductive AFM measurements were done with an ICON (Bruker) microscope operated in an air-conditioned laboratory ($T_{amb}$ = 22.5 °C, relative humidity of 35-40 %). The AFM images were treated with the Gwyddion software (http://gwyddion.net/). Topographic images were acquired in tapping mode (TM) using a silicon tip (42 N/m spring constant, resonance frequency 320 kHz).

To determine the perpendicular conductivity, the current–voltage characteristics were measured by conductive atomic force microscopy (C-AFM) using a PtIr coated tip (SCM-PIT-V2 from Bruker, 3.0 N/m spring constant). We first focused on the middle of a nanowire by imaging a small zone 40x40 nm² in the TM mode with the C-AFM tip. Then we switched to a stationary C-AFM mode on the center of the NWs (no scan) to locally measure the I-V curve at a loading force of ~ 60nN. Around 10 I-V curves were measured on each nanowire. The measurements were repeated on 4 different NWs. The voltage was applied on the substrate, and the tip was grounded via the input of the transimpedance preamplifier.

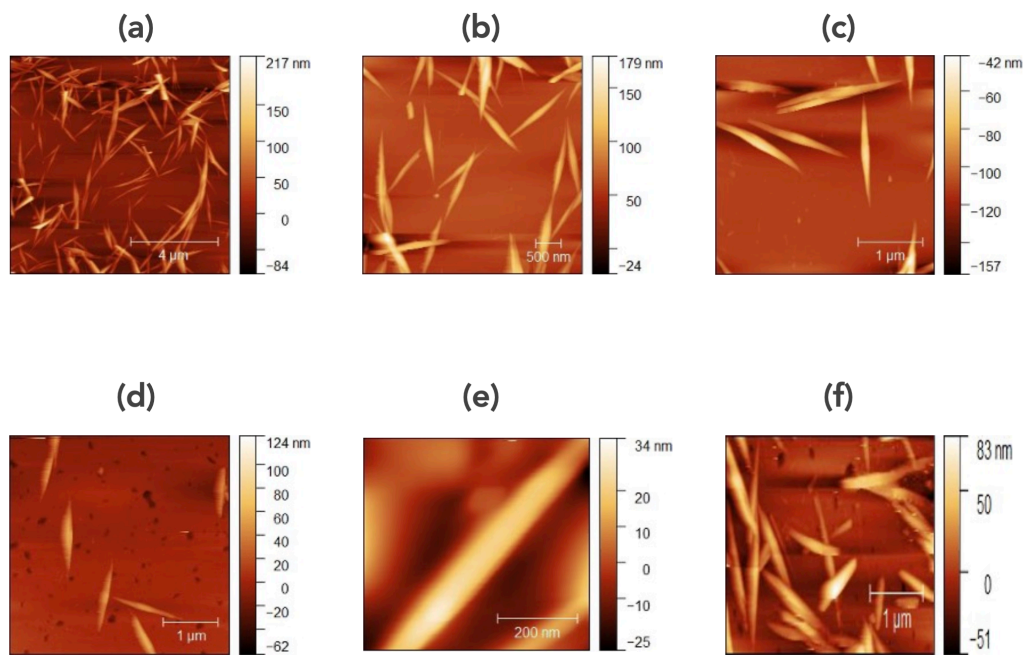

**Figure S1**. Additional TM-AFM images of the NWs on SiO$_2$ (a) 10 μm x 10 μm, (b) and (c) 3 μm x 3 μm, 2 different zones from the area shown in the panel (a). (d-e) TM-AFM images of the NWs on template-stripped Au surface at various magnifications: 4 μm x 4 μm, 0.6 μm x 0.6 μm and 4 μm x 4 μm, respectively.

Ultraflat template-stripped gold surfaces ($^{TS}$Au), with a rms roughness of ∼0.4 nm, were prepared according to methods already reported.[1-3] In brief, a 300–500 nm thick Au film was evaporated on a very flat silicon wafer covered by its native SiO$_2$ (rms roughness of ∼0.4 nm), which was previously carefully cleaned by piranha solution (30 min in 7:3 H$_2$SO$_4$/H$_2$O$_2$ (v/v); **Caution**: Piranha solution is a strong oxidizer and reacts exothermically with organics), rinsed with deionized (DI) water, and dried under a stream of nitrogen. Clean 10x10 mm pieces of glass slide (ultrasonicated in acetone for 5 min, ultrasonicated in 2-propanol for 5 min, and UV irradiated in ozone for 10 min) were glued on the evaporated Au film (UV-polymerizable glue, NOA61 from Epotecny), then mechanically peeled off providing the $^{TS}$Au film attached on the glass side.

### S3. TEM and HR-TEM characterization.

Electron microscopy characterizations were carried out on three microscopes:

- Jeol 1011 operated at a voltage of 100 kV and enabling morphological analysis at magnifications < 100k.

- Jeol 2100Plus (LaB6 emission) and Jeol 2100 F (Field effect emission) operated both at a voltage of 200 kV, with a lattice resolution of 1.4 Å, enabling high-resolution imaging.

A small amount of powder was diluted in purged ethanol and then deposited in a carbon grid by drop casting. The images were acquired thanks to a Gatan Orius camera (on Jeol 1011 and 2100Plus) and Gatan US400 camera (on Jeol 2100 F). Twenty-eight NWs were imaged (as in Fig. 2 in the main text and Fig. S2) and analyzed using ImageJ (https://imagej.nih.gov/ij/index.html) to extract the mean values of the external diameter ($d_{nt}$) and the internal one ($\delta_{nt}$). All the measured values are given in the table S1 below.

| # NT | $\delta_{nt}$ (nm) | $d_{nt}$ (nm) | # NT | $\delta_{nt}$ (nm) | $d_{nt}$ (nm) |
|---|---|---|---|---|---|
| #1 | 1.0 | 3.1 | #17 | 0.9 | 2.9 |
| #2 | 1.1 | 2.5 | #18 | 1.0 | 2.1 |
| #3 | 0.7 | 2.5 | #19 | 1.0 | 2.3 |
| #4 | 0.9 | 2.0 | #20 | 1.1 | 2.8 |
| #5 | 0.9 | 2.4 | #21 | 1.0 | 2.5 |
| #6 | 0.8 | 2.5 | #22 | 0.9 | 2.9 |
| #7 | 2.2 | 4.1 | #23 | 0.8 | 2.5 |
| #8 | 1.8 | 3.5 | #24 | 0.9 | 2.7 |
| #9 | 1.2 | 3.7 | #25 | 0.9 | 2.8 |
| #10 | 1.7 | 5.3 | #26 | 0.6 | 2.0 |
| #11 | 1.1 | 2.7 | #27 | 0.9 | 2.8 |
| #12 | 0.9 | 2.7 | #28 | 0.9 | 2.2 |
| #13 | 0.6 | 2.2 | Max | 2.2 | 5.3 |
| #14 | 0.7 | 2.2 | Min | 0.6 | 2 |
| #15 | 0.9 | 2.9 | Mean | 1.0 ± 0.35 | 2.7 ± 0.7 |
| #16 | 1.0 | 2.3 | | | |

**Table S1.** *External diameter ($d_{nt}$) and internal diameter ($\delta_{nt}$) measured on 28 TEM images.*

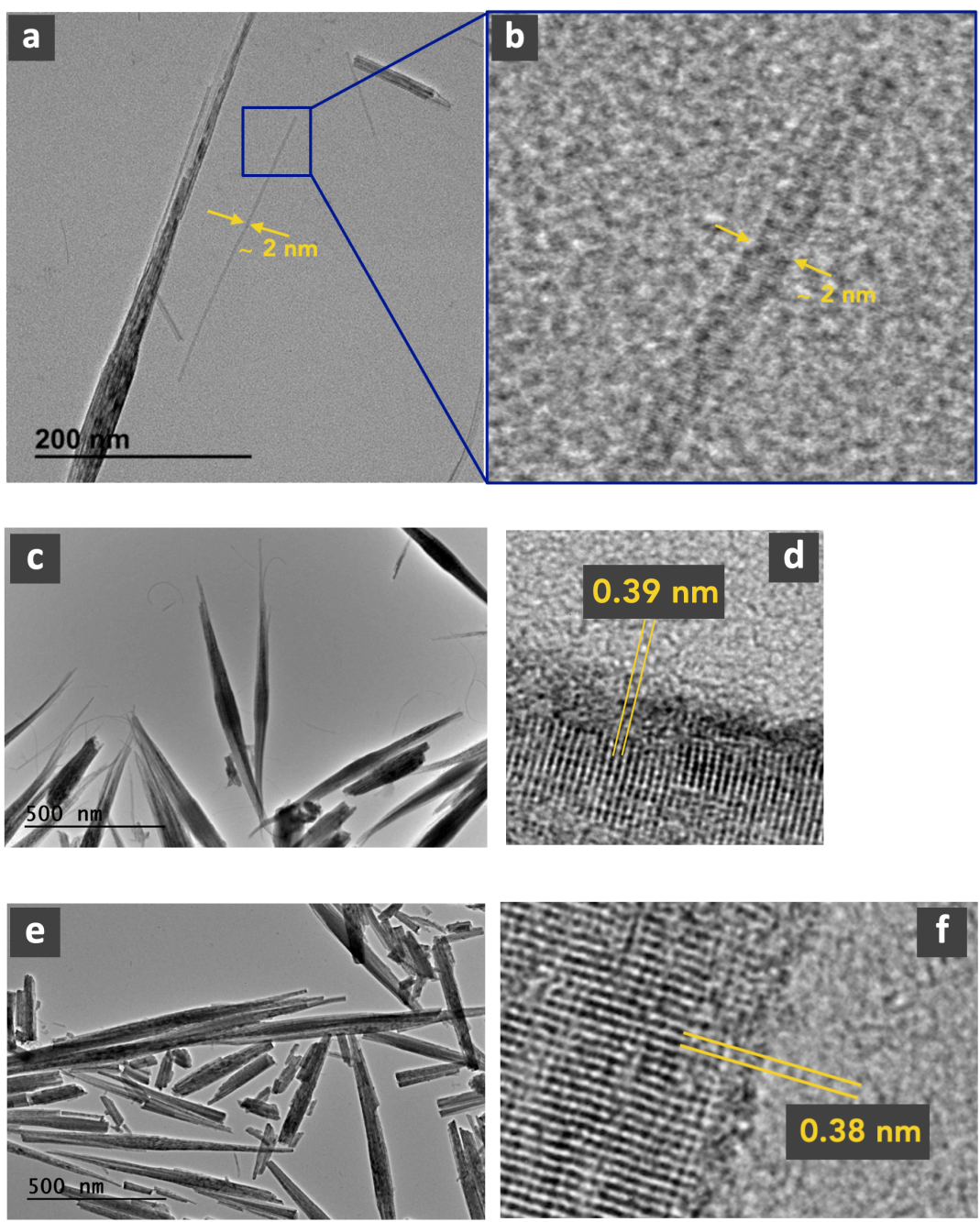

*Figure S2*. TEM and HR-TEM images. (a) TEM image showing a single NT (selected from Fig. 2d in the main text) and (b) zoom of the area framed in blue allowing distinguishing the reticular planes. (c-d) TEM and HR-TEM images of freshly prepared samples. (e-f) TEM and HR-TEM images of an aged sample (19 months).

| #NT | d (Å) FFT | d (Å) profile |
|---|---|---|
| #1 | 3.67 | 3.71 |
| #2 | 3.88 | 3.83 |
| #3 | 3.84 | 3.88 |
| #4 | 3.67 | 3.78 |
| #5 | 3.72 | 3.73 |
| #6 | 3.74 | 3.73 |
| #7 | 3.82 | 3.84 |
| #8 | 3.88 | 3.86 |
| #9 | 3.89 | 3.90 |
| #10 | 3.80 | 3.87 |

**Table S2.** *Lattice parameters along the growth direction measured on 10 NTs by HR-TEM from the FFT pattern (as in the inset of Fig. 2f in the main text) and directly from the fringes on the HR-TEM images.*

## S4. X-ray powder diffraction (XRD) and Raman spectroscopy measurements.

*XRD.* We used a Rigaku MM007HF diffractometer equipped with Varimax focusing optics, a RAXIS4++ image plate detector and a Mo rotating anode ($\lambda K_{\alpha 1}$ = 0.709319 Å and $\lambda K_{\alpha 2}$ = 0.713609 Å) at 50 keV and 24 mA. The samples were placed in a 0.25 mm cryoloop, data were collected with a scan range 2Θ from 3 to 45° and an acquisition time of 10 minutes in the transmission geometry. The Fit2D program[4] was used for the azimuthal integration of 2D images into 1D patterns (from 3 to 45°) after a calibration with a LaB6 standard. Le Bail refinements[5] were performed with the FullProf suite of programs.[6] The starting unit cell parameters were taken from ICSD-15254 and ICSD 1620,[7, 8] for the crystal structure of $W_{18}O_{49}$ and the triclinic phase of $WO_3$, respectively. The large unit cell parameters and/or the low symmetry of $W_{18}O_{49}$ and $WO_3$ compounds lead to a large number of Bragg peak positions, which do not allow to unambiguously determine which crystal structure is the correct one. It is possible to equally fit the same experimental data with both starting models. Figure S3 shows the same experimental data as in Fig. 3a but fitted with a WO$_3$ structure (ICDS 1620) instead of W$_{18}$O$_{49}$ (WO$_{2.72}$) model as in Fig. 3a.

| Starting model | W$_{18}$O$_{49}$ (ICSD 15524) | WO$_3$ (ICSD 1620) |
|---|---|---|
| Crystal system | Monoclinic | Triclinic |
| Space group | *P2/m* | *P-1* |
| *a* [Å] | 18.321(15) | 7.158(10) |
| *b* [Å] | 3.796(2) | 7.524(4) |
| *c* [Å] | 14.011(17) | 7.650(2) |
| *α* [°] | 90 | 87.23(6) |
| *β* [°] | 115.32(6) | 89.48(9) |
| *γ* [°] | 90 | 92.51(9) |
| $R_p = \frac{\sum |y_o - y_c|}{\sum y_o}$ | 1.12 | 0.83 |
| $\omega R_p = \sqrt{\frac{\sum \left(\omega (y_o^2 - y_c^2)^2\right)}{\sum \left(\omega (y_o^2)^2\right)}}$ | 2.37 | 1.18 |

***Table S3.*** *Refinement data table: unit cell parameters and reliability factors.*

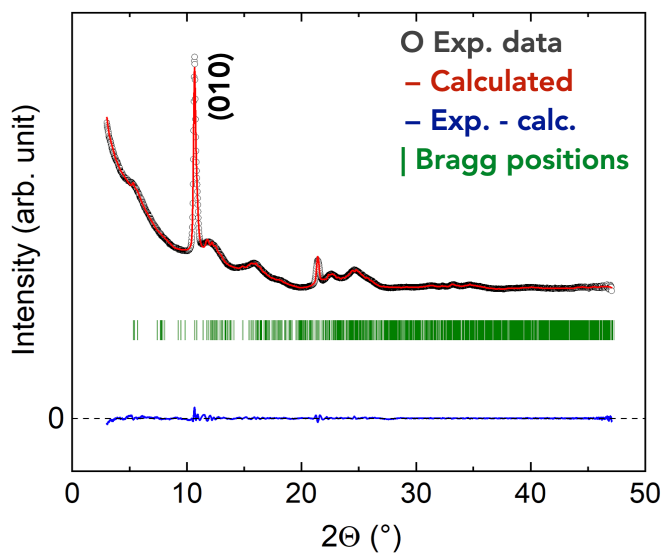

***Figure S3****. XRD patterns (Mo-Kα radiation): the measured pattern (black open circles, same data as in Fig. 3a), the calculated pattern (red line), and difference between the two (blue line). The green bars show the expected positions of Bragg peaks related to WO$_3$ structure (ICDS 1620 crystal structure data).*

*Raman spectroscopy.* We used a LabRAM HR confocal system from Horiba Jobin-Yvon. We used a 473 nm excitation laser (≈1 mW) focused with a 100× objective for the confocal Raman spectroscopy measurements in air at room temperature. We used a 1800 grooves per mm grating, a spot size of ∼1 μm and a resolution of 1 cm$^{-1}$. A few mg of NWs were deposited on a glass substrate. On each sample, 3 measurements were performed at different locations on the sample and then averaged. Raman data were treated with Labspec5 software from Bruker.

## S5. XPS measurements.

High resolution XPS spectra were recorded with a Physical Electronics (PHI) type 5600 spectrometer. We used a monochromatic Al$_{K\alpha}$ X-ray source (hυ = 1486.6 eV), a detection angle of 45° as referenced to the sample surface, an analyzer entrance slit width of 400 μm and an analyzer pass energy of 12 eV. In these conditions, the overall resolution as measured from the full-width half-maximum (FWHM) of the Ag 3d$_{5/2}$ line is 0.55 eV. Alternatively, XPS analyses were performed using an Omicron Argus X-ray photoelectron spectrometer. The emission of photoelectrons from the sample was analyzed at a takeoff angle of 45° under ultra-high vacuum conditions (≤ 10$^{-10}$ Torr). The spectra were acquired with a 20 eV pass energy. The XPS spectra were fitted using the PHI multipak software or the casaXPS software.[9] The peaks were decomposed using Voigt functions and a least squares minimization procedure. Binding energies were referenced to the C 1s binding energy, set at 284.8 eV. The amplitude ratio between the 4f$_{7/2}$ and 4f$_{5/2}$ peaks is fixed at 4/3, the energy splitting between the two peaks is let adjustable, the values were found ≈ 2.1-2.3 eV, in good agreement with the reported values of 2.2 eV.[10] Table S4 summarizes the fitted parameters (peak position, integrated peak area and FWHM) for the W 4f and W 5p peaks of the three measured samples (Fig. 3c and Fig. S4). The FWHM of the W$^{6+}$ and W$^{5+}$ 4f peaks are 1.6-1.9 eV and 1.1-1.2 eV, respectively, in agreement with reported results.[11-14] The 5p$_{3/2}$ peaks are broader as also reported.[15]

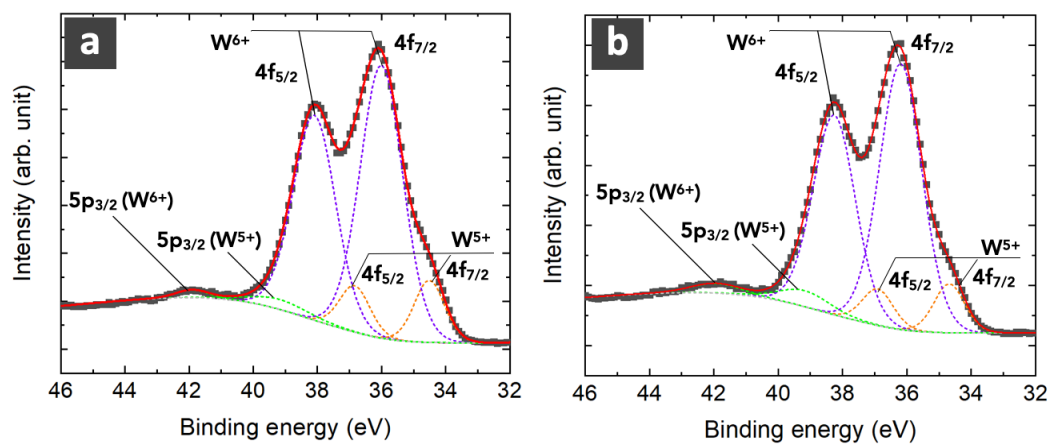

**Figure S4.** XPS spectra (W 4f band) of two freshly prepared samples (two batches processed with the same solvothermal conditions) with the peak deconvolution giving a stoichiometry $WO_{2.93}$ and $WO_{2.94}$, respectively (experimental data: black squares; fit: red line; deconvolution: dashed lines).

|  | sample (Fig. 3c) | | | sample (Fig. S4a) | | | sample (Fig. S4b) | | |
| --- | --- | --- | --- | --- | --- | --- | --- | --- | --- |
|  | poisition (eV) | area (arb. unit) | FWHM (eV) | poisition (eV) | area (arb. unit) | FWHM (eV) | poisition (eV) | area (arb. unit) | FWHM (eV) |
| $W^{6+}$ $4f_{5/2}$ | 38.3 | 3776 | 1.90 | 38.3 | 10787 | 1.56 | 38.1 | 14028 | 1.58 |
| $W^{6+}$ $4f_{7/2}$ | 36.1 | 4977 | 1.90 | 36.2 | 14383 | 1.56 | 36.0 | 18704 | 1.58 |
| $W^{5+}$ $4f_{5/2}$ | 36.8 | 2839 | 1.17 | 36.9 | 1429 | 1.13 | 36.8 | 2264 | 1.12 |
| $W^{5+}$ $4f_{7/2}$ | 34.6 | 3786 | 1.17 | 34.7 | 1906 | 1.13 | 34.6 | 3018 | 1.12 |
| $W^{6+}$ $5p_{3/2}$ | 40.9 | 557 | 3.7 | 41.9 | 672 | 1.83 | 41.9 | 440 | 1.20 |
| $W^{5+}$ $5p_{3/2}$ | 39.4 | 349 | 3.9 | 39.2 | 1158 | 1.87 | 39.2 | 1026 | 2.13 |

**Table S4.** Fitted parameters of the XPS W 4f and 5p bands for the three samples shown in Fig. 3c and Fig. S4.

## S6. 4-probe STM in UHV.

The longitudinal conductivity of the NWs was measured in ultrahigh vacuum (UHV, ≤ $10^{-10}$ Torr) with a multiple-probe STM (Nanoprobe, Scienta-Omicron) equipped with four independent STM scanners for imaging and contacting nanostructures. The 4P-STM is equipped with a scanning electron microscope (SEM) for monitoring the tip position. We used Nanonis STM controllers to operate the STM probes at the nanoscale with Keithley Source-Measure Units (SMU) to perform the electrical measurements. The tungsten tips were prepared by an electrochemical etching in NaOH and thoroughly annealed in the UHV preparation chamber to remove the thin oxide layer covering the tips. The NWs were deposited on a 200 nm thick $SiO_2$/Si sample and the STM tips were approached under the supervision of the SEM. As the $SiO_2$ surface is insulating, the usual STM tip-surface distance control system, based on the tunnel current measurement, could not be used for the approach of the first tip. A high bias was applied on the tip (typically -8 V) and it was manually approached close to the NW until a charging hollow effect of the $SiO_2$ was visible on the SEM image around the NW. After setting the bias back to 0 V on the first tip kept in contact with the NW, the approach of the three other STM tips was controlled by monitoring the tunneling current between the tip and the NW. The contact was detected at the crossover of the exponential distance-dependent tunnel current and the weak distance-dependent contact regime. For the measurements below 300K, the tips were retracted far enough from the sample during the cooling down to avoid any damage of the NW that could be induced by the thermal drifts of the sample holder and the tips. Once the temperature was reached, we waited half an hour to stabilize the system before contacting the NWs as described above and start again the measurements.

*Approximation to calculate the longitudinal conductivity*. Since there is no analytical formula to calculate the longitudinal conductivity from the measured

S10

conductance $G_L$ for a spindle-shaped NW, we used the classical equation $\sigma_{Lnw}=G_L(L/S_{nw})$, where L is the inner-probe distance (L) and $S_{nw}$ the cross-section area of a cylinder-shaped NW. We estimated the induced error by considering two cases. In case 1, we used the cross-section area of the NW, $S_{nw}$, estimated by measuring the NW diameter (from SEM image) at its thickest part in the middle. In case 2, $S_{nw}$ was estimated taking the average diameter between the values at its thickest part and at the inner-probe contacts (Fig. S5), which are almost similar at the two probe positions. We used the $G_L$ value measured on the same NW with different inner-probe distances L (data of NW #1 of the second series in Table S5). The difference of the calculated conductivity (table in Fig. S5) is clearly below the data dispersion (Fig. 4c main text) observed for the complete dataset, and in both cases, the calculated values are around the max of the log-normal distribution. Thus, for simplicity, we used the method of case 1 throughout this work to calculate the longitudinal conductivity reported in Tables S5, S6 and shown in Fig. 4.

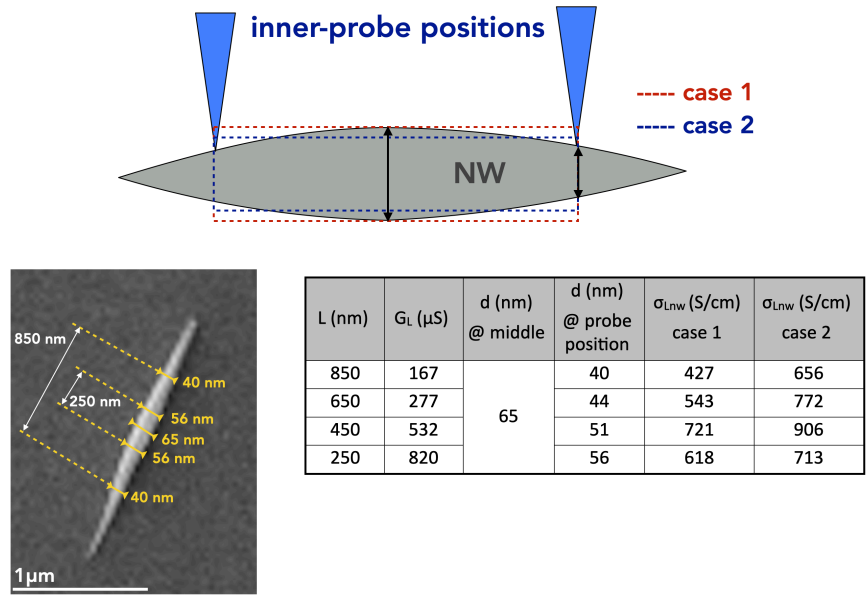

| L (nm) | $G_L$ (µS) | d (nm) @ middle | d (nm) @ probe position | $\sigma_{Lnw}$ (S/cm) case 1 | $\sigma_{Lnw}$ (S/cm) case 2 |
|---|---|---|---|---|---|
| 850 | 167 | 65 | 40 | 427 | 656 |
| 650 | 277 |  | 44 | 543 | 772 |
| 450 | 532 |  | 51 | 721 | 906 |
| 250 | 820 |  | 56 | 618 | 713 |

*Figure S5. Scheme of the NW and the two inner probes. The dashed lines indicate the approximation using a cylinder-shaped structure to calculate the longitudinal*

*conductivity. The dimensions were evaluated from the SEM image of the NW (here acquired just before the deposition of the probes). The yellow arrows indicate the measured diameters at the middle (the thickest part of the NW) and at the location of the two inner probes. For clarity, only two inner-probe distances (850 and 250 nm) are shown. The table summarizes the calculated longitudinal conductivity in the two approximation cases.*

| #NW 1ts series | $G_L$ (µS) | L (nm) | $d_{nw}$ (nm) | $S_{nw}$ (nm²) | $\sigma_{Lnw}$ (S/cm) | N | $\sigma_{Lnt}$ (S/cm) |
|---|---|---|---|---|---|---|---|
| #1 | 741 | 175 | 70 | 3848 | 337 | 672 | 390 |
| #2 | 500 | 245 | 46 | 1662 | 737 | 290 | 854 |
| #3 | 455 | 400 | 75 | 4418 | 412 | 772 | 477 |
| #4 | 303 | 300 | 50 | 1963 | 463 | 343 | 537 |
| #5 | 370 | 350 | 50 | 1963 | 660 | 343 | 765 |
| #6 | 444 | 500 | 80 | 5027 | 442 | 878 | 512 |
| #7 | 345 | 660 | 75 | 4418 | 515 | 772 | 597 |
| #8 | 370 | 300 | 60 | 2827 | 393 | 494 | 455 |
| #9 | 741 | 310 | 66 | 3421 | 671 | 598 | 778 |
| #10 | 286 | 640 | 66 | 3421 | 534 | 598 | 619 |
| #10 | 588 | 420 | 66 | 3421 | 722 | 598 | 837 |
| #10 | 118 | 230 | 66 | 3421 | 791 | 598 | 917 |
| #11 | 320 | 730 | 77 | 4657 | 502 | 813 | 581 |

| #NW 2nd series | $G_L$ (µS) | L (nm) | $d_{nw}$ (nm) | $S_{nw}$ (nm²) | $\sigma_{Lnw}$ (S/cm) | N | $\sigma_{Lnt}$ (S/cm) |
|---|---|---|---|---|---|---|---|
| #1 | 167 | 850 | 65 | 3318 | 427 | 580 | 495 |
| #1 | 277 | 650 | 65 | 3318 | 543 | 580 | 629 |
| #1 | 532 | 450 | 65 | 3318 | 721 | 580 | 836 |
| #1 | 820 | 250 | 65 | 3318 | 618 | 580 | 716 |
| #3 | 226 | 574 | 64 | 3217 | 403 | 562 | 467 |
| #3 | 346 | 400 | 64 | 3217 | 430 | 562 | 499 |
| #3 | 455 | 200 | 64 | 3217 | 283 | 562 | 328 |
| #5 | 221 | 725 | 67 | 3526 | 454 | 616 | 526 |
| #5 | 588 | 500 | 67 | 3526 | 834 | 616 | 967 |
| #5 | 746 | 300 | 67 | 3526 | 635 | 616 | 736 |
| #5 | 116 | 150 | 67 | 3526 | 495 | 616 | 573 |
| #6 | 40 | 816 | 45 | 1590 | 207 | 278 | 240 |
| #6 | 54 | 600 | 45 | 1590 | 203 | 278 | 236 |
| #6 | 61 | 400 | 45 | 1590 | 152 | 278 | 177 |
| #6 | 55 | 200 | 45 | 1590 | 69 | 278 | 80 |
| #7 | 301 | 412 | 60 | 2827 | 439 | 494 | 509 |
| #7 | 318 | 300 | 60 | 2827 | 338 | 494 | 392 |
| #7 | 345 | 200 | 60 | 2827 | 244 | 494 | 283 |
| #8 | 173 | 490 | 60 | 2827 | 300 | 494 | 348 |
| #8 | 183 | 350 | 60 | 2827 | 226 | 494 | 262 |
| #8 | 194 | 200 | 60 | 2827 | 137 | 494 | 159 |
| #9 | 385 | 528 | 55 | 2376 | 855 | 415 | 991 |
| #9 | 556 | 300 | 55 | 2376 | 702 | 415 | 813 |
| #9 | 855 | 150 | 55 | 2376 | 540 | 415 | 625 |
| #10 | 315 | 595 | 68 | 3632 | 517 | 634 | 599 |
| #10 | 417 | 400 | 68 | 3632 | 459 | 634 | 532 |
| #10 | 585 | 200 | 68 | 3632 | 322 | 634 | 373 |
| #11 | 192 | 550 | 60 | 2827 | 374 | 494 | 434 |
| #11 | 224 | 400 | 60 | 2827 | 316 | 494 | 367 |
| #11 | 380 | 200 | 60 | 2827 | 269 | 494 | 312 |

| #NW 3rd series | $G_L$ (µS) | L (nm) | $d_{nw}$ (nm) | $S_{nw}$ (nm²) | $\sigma_{Lnw}$ (S/cm) | N | $\sigma_{Lnt}$ (S/cm) |
|---|---|---|---|---|---|---|---|
| #1  | 442 | 380 | 72 | 4072 | 413 | 711  | 479 |
| #2  | 108 | 500 | 66 | 3421 | 157 | 598  | 182 |
| #3  | 415 | 450 | 64 | 3217 | 580 | 562  | 673 |
| #4  | 67  | 574 | 58 | 2642 | 147 | 461  | 170 |
| #15 | 365 | 619 | 96 | 7238 | 312 | 1264 | 362 |
| #16 | 93  | 428 | 61 | 2922 | 137 | 510  | 158 |
| #17 | 181 | 624 | 80 | 5027 | 225 | 878  | 261 |
| #18 | 223 | 474 | 81 | 5153 | 205 | 900  | 237 |
| #19 | 105 | 528 | 86 | 5809 | 96  | 1015 | 111 |

*Table S5.* The complete dataset at 300 K: zero-bias NW longitudinal conductance ($G_L$) from data in Fig. 4b (main text), inner-probe distance (L), NW diameter and surface ($d_{nw}$ and $S_{nw}$), longitudinal NW conductance ($\sigma_{Lnw}=G_L(L/S_{nw})$), number of NTs in the NW ($N \approx (d_{nw}/d_{nt})^2$ with $d_{nt}=2.7$ nm (± 0.7 nm), see main text) and longitudinal individual NT conductivity ($\sigma_{Lnt}=G_L(L/S_{nt})/N$ with $S_{nt}$ the cross-section surface of the NT, see main text).

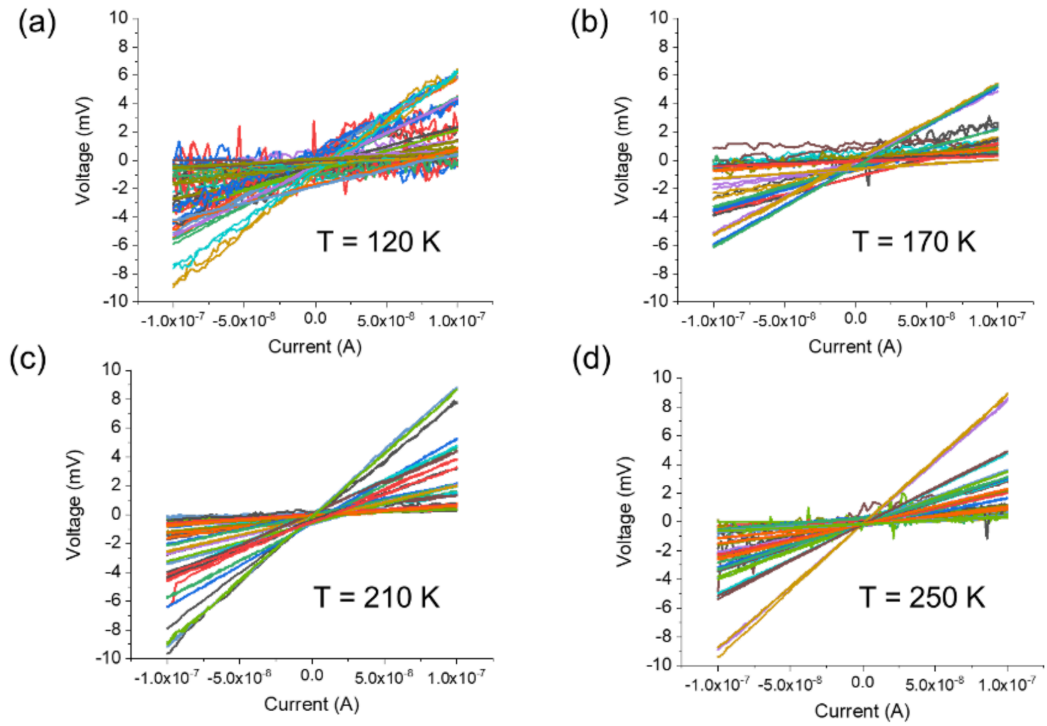

*Figure S6.* V-I datasets at T=120K, 170K, 210K, 250K.

| #NW @120K | $G_L$ (μS) | L (nm) | $d_{nw}$ (nm) | $S_{nw}$ (nm²) | $\sigma_{Lnw}$ (S/cm) | N | $\sigma_{Lnt}$ (S/cm) |
|---|---|---|---|---|---|---|---|
| #1 | 19 | 411 | 72 | 4072 | 19 | 711 | 23 |
| #1 | 15 | 417 | 72 | 4072 | 15 | 711 | 17 |
| #1 | 277 | 417 | 72 | 4072 | 284 | 711 | 329 |
| #3 | 78 | 440 | 64 | 3217 | 106 | 562 | 123 |
| #4 | 76 | 283 | 58 | 2642 | 82 | 461 | 95 |
| #7 | 238 | 503 | 80 | 5027 | 238 | 878 | 275 |
| #9 | 19 | 480 | 60 | 2827 | 32 | 494 | 37 |
| #20 | 26 | 615 | 75 | 4418 | 37 | 772 | 42 |
| #21 | 20 | 350 | 55 | 2376 | 29 | 415 | 34 |
| #22 | 13 | 470 | 57 | 2552 | 24 | 446 | 28 |
| #23 | 77 | 638 | 69 | 3739 | 132 | 653 | 153 |
| #24 | 10 | 490 | 66 | 3421 | 15 | 598 | 17 |
| #25 | 155 | 496 | 62 | 3019 | 255 | 527 | 295 |
| #26 | 39 | 505 | 73 | 4185 | 47 | 731 | 55 |
| #27 | 121 | 673 | 62 | 3019 | 270 | 527 | 313 |
| #28 | 232 | 520 | 62 | 3019 | 400 | 527 | 463 |
| #29 | 48 | 420 | 83 | 5411 | 37 | 945 | 43 |
| #30 | 18 | 522 | 79 | 4902 | 19 | 856 | 22 |
| #31 | 36 | 562 | 65 | 3318 | 61 | 580 | 70 |

| #NW @170K | $G_L$ (μS) | L (nm) | $d_{nw}$ (nm) | $S_{nw}$ (nm²) | $\sigma_{Lnw}$ (S/cm) | N | $\sigma_{Lnt}$ (S/cm) |
|---|---|---|---|---|---|---|---|
| #3 | 37 | 350 | 64 | 3217 | 40 | 562 | 47 |
| #7 | 187 | 419 | 78 | 4778 | 164 | 835 | 190 |
| #9 | 54 | 346 | 60 | 2827 | 67 | 494 | 77 |
| #10 | 145 | 516 | 56 | 2463 | 304 | 430 | 352 |
| #10 | 130 | 516 | 56 | 2463 | 272 | 430 | 315 |
| #10 | 60 | 516 | 56 | 2463 | 126 | 430 | 146 |
| #32 | 37 | 393 | 75 | 4418 | 33 | 772 | 38 |
| #33 | 35 | 328 | 61 | 2922 | 39 | 510 | 45 |
| #34 | 20 | 487 | 73 | 4185 | 23 | 731 | 27 |
| #35 | 119 | 585 | 56 | 2463 | 282 | 430 | 327 |
| #36 | 233 | 479 | 63 | 3117 | 357 | 544 | 414 |
| #37 | 18 | 433 | 66 | 3421 | 23 | 598 | 26 |
| #38 | 147 | 392 | 76 | 4536 | 127 | 792 | 147 |

| #NW @210K | $G_L$ (μS) | L (nm) | $d_{nw}$ (nm) | $S_{nw}$ (nm²) | $\sigma_{Lnw}$ (S/cm) | N | $\sigma_{Lnt}$ (S/cm) |
|---|---|---|---|---|---|---|---|
| #2 | 23 | 321 | 66 | 3421 | 22 | 598 | 26 |
| #7 | 280 | 250 | 78 | 4778 | 147 | 835 | 170 |
| #9 | 158 | 272 | 60 | 2827 | 152 | 494 | 176 |
| #10 | 67 | 237 | 56 | 2463 | 65 | 430 | 75 |
| #11 | 68 | 658 | 57 | 2552 | 177 | 446 | 205 |
| #12 | 96 | 492 | 57 | 2552 | 185 | 446 | 215 |
| #13 | 244 | 487 | 66 | 3421 | 347 | 598 | 402 |
| #39 | 165 | 508 | 68 | 3632 | 230 | 634 | 267 |
| #41 | 43 | 378 | 57 | 2552 | 64 | 446 | 74 |
| #42 | 36 | 397 | 51 | 2043 | 70 | 357 | 82 |
| #43 | 148 | 517 | 68 | 3632 | 211 | 634 | 245 |
| #44 | 47 | 524 | 48 | 1810 | 136 | 316 | 158 |
| #45 | 42 | 629 | 66 | 3421 | 77 | 598 | 89 |
| #46 | 22 | 506 | 68 | 3632 | 31 | 634 | 36 |
| #47 | 154 | 637 | 75 | 4418 | 222 | 772 | 257 |
| #49 | 27 | 688 | 58 | 2642 | 69 | 461 | 80 |

| #NW @250K | $G_L$ (μS) | L (nm) | $d_{nw}$ (nm) | $S_{nw}$ (nm²) | $\sigma_{Lnw}$ (S/cm) | N | $\sigma_{Lnt}$ (S/cm) |
|---|---|---|---|---|---|---|---|
| #7 | 198 | 312 | 78 | 4778 | 129 | 835 | 150 |
| #9 | 128 | 221 | 60 | 2827 | 100 | 494 | 116 |
| #10 | 169 | 212 | 56 | 2463 | 145 | 430 | 168 |
| #11 | 262 | 285 | 57 | 2552 | 293 | 446 | 340 |
| #13 | 162 | 556 | 66 | 3421 | 263 | 598 | 305 |
| #13 | 127 | 598 | 66 | 3421 | 222 | 598 | 257 |

| #NW @250K | $G_L$ (µS) | L (nm) | $d_{nw}$ (nm) | $S_{nw}$ (nm²) | $\sigma_{Lnw}$ (S/cm) | N | $\sigma_{Lnt}$ (S/cm) |
|---|---|---|---|---|---|---|---|
| #14 | 120 | 570 | 77 | 4657 | 147 | 813 | 170 |
| #50 | 60 | 701 | 61 | 2922 | 144 | 510 | 167 |
| #51 | 100 | 619 | 72 | 4072 | 153 | 711 | 177 |
| #52 | 195 | 481 | 66 | 3421 | 274 | 598 | 318 |
| #53 | 41 | 580 | 69 | 3739 | 64 | 653 | 74 |
| #54 | 32 | 599 | 70 | 3848 | 50 | 672 | 58 |
| #55 | 73 | 574 | 71 | 3959 | 105 | 691 | 122 |
| #56 | 27 | 663 | 66 | 3421 | 52 | 598 | 60 |
| #57 | 42 | 646 | 72 | 4072 | 66 | 711 | 77 |
| #58 | 32 | 629 | 63 | 3117 | 64 | 544 | 75 |
| #60 | 20 | 562 | 62 | 3019 | 38 | 527 | 44 |
| #61 | 42 | 728 | 71 | 3959 | 77 | 691 | 89 |
| #62 | 201 | 588 | 76 | 4536 | 261 | 792 | 302 |

*Table S6.* Complete datasets at 120, 170, 210 and 250 K: zero-bias NW conductance ($G_L$) from data in Fig. S6, inner-probe distance (L), NW diameter and surface ($d_{nw}$ and $S_{nw}$), longitudinal NW conductance ($\sigma_{Lnw}=G_L(L/S_{nw})$), number of NTs in the NW ($N \approx (d_{nw}/d_{nt})^2$ with $d_{nt}$=2.7 nm (± 0.7 nm), see main text) and longitudinal individual NT conductivity ($\sigma_{Lnt}=G_L(L/S_{nt})/N$ with $S_{nt}$ the cross-section surface of the NT, see main text).

## S7. Effect of e-beam exposure.

The NWs were imaged by the SEM embodied in the 4-probe STM machine to visualize them for a precise positioning of the 4 STM tips. We evaluated a possible influence of the e-beam exposure on their electrical conductivity. To do so, we fabricated a 2-dimensional (2D) percolated network of NWs on lithographed Au electrodes (Fig. S7a). The electrodes were fabricated on a Si/SiO$_2$ (200 nm thick) substrate by a standard lithography process. The electrodes, Ti (2 nm, adhesion layer) and Au (12 nm), are 1 mm long and spaced by 5 µm. The NWs were deposited by drop casting. Figure S7b shows the I-V curve measured on the same 2D networks before and after exposure to the e-beam of the SEM in the same conditions as in the 4-probe STM experiment (e-beam at 10 kV during ~ 20 min). To prevent any influence of the ambient air, the I-Vs were measured in a glove box (under dry N$_2$, < 1 ppm of oxygen and water vapor). We note a slight increase of the current by a factor ~ 1.5. Such a correction factor was taken into account to compare the longitudinal conductivity measured by the 4P-STM with e-beam (SEM) exposure (see section S6) and the perpendicular conductivity, this latter being measured by C-AFM (see section S2) without exposition to the e-beam of the SEM.

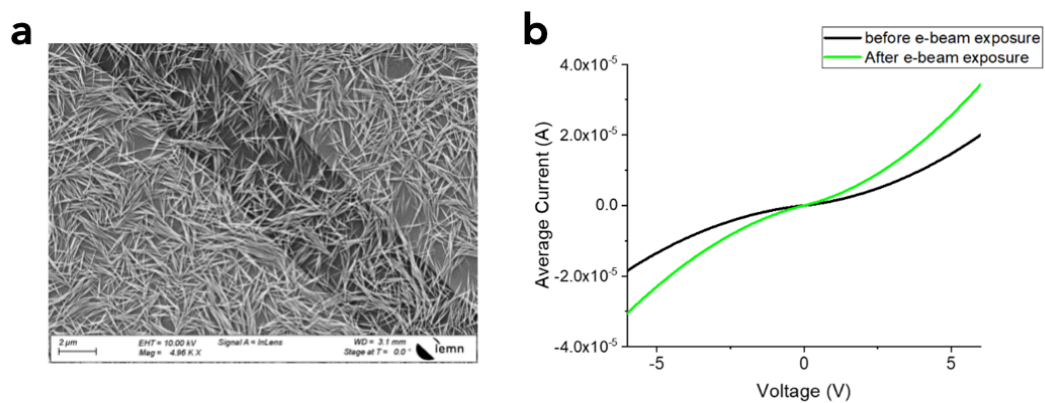

*Figure S7*. SEM image of a percolating 2D network of NWs between two Au electrodes. Average I-V (from measurements on 5 different 2D networks) before (dark line) and after (green line) e-beam exposure.

## S8. Other electron transport mechanism.

We tested a classical temperature-activated transport mechanism for the samples with the lowest conductivity (data from Fig. 5b, see main text). Figure S8a shows the Arrhenius plot (ln($\sigma_L$) vs. 1000/T) of the same data as in Fig. 5b. These data badly follow an Arrhenius plot, moreover, the liner fit of this plot gives a physically insignificant low value of the activation energy of 26 ± 4 meV. We also tested a polaron hopping transport as suggested for $W_{18}O_{49}$ nanowires synthesized by a low temperature (600°C) furnace process.[16] Figure S8b shows the plot of ln($\sigma_L T$) vs. 1000/T with the same data. Again, we obtained a poor linear behavior with a low activation energy (42 ± 6 meV) compared to ~0.25 eV as previously reported [16].

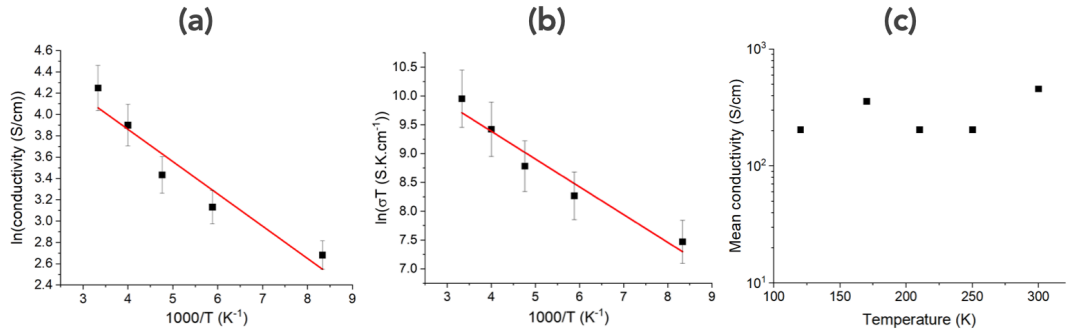

***Figure S8***. *(a) Plot of ln($\sigma_L$) vs. 1000/T and (b) plot of ln($\sigma_L T$) vs. 1000/T with the same data as in Fig. 5b. The red lines are the fits by a linear regression (bad fits with $r^2$ = 0.92 and 0.94, respectively). (c) Mean longitudinal conductivity (calculated from the statistical distributions shown in Fig. 5a, main text).*

## S9. Transistor configuration.

We measured at RT the current-voltage characteristic of NW connected by two STM probes and we applied a gate voltage on the underneath highly doped Si substrate. We did not observe any modulation of the current in the NW with the applied gate voltage (no field effect contrary to the expectations for semiconducting NWs).

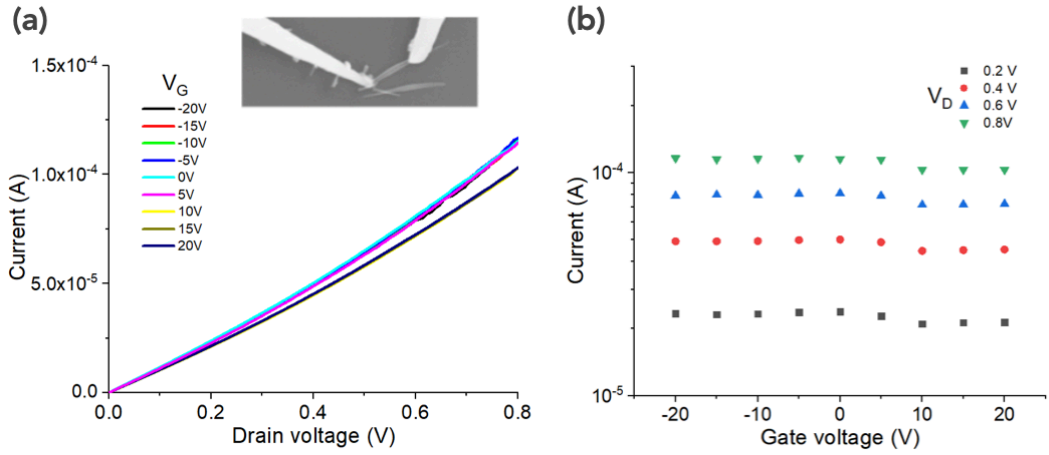

***Figure S9.*** *(a) Drain current vs. drain voltage (2 probes) for various gate voltages (transistor configuration, -20V < $V_G$ < 20 V). (b) Same data plotted as a transfer characteristic: drain voltage vs. gate voltage at several drain voltages. The slight difference between the curves at different gate voltages is not significant and it is likely due to a small shift of the probe positions along the NW or small variations of the contact resistance during the measurements.*

## S10. The C-AFM dataset, perpendicular conductivity.

| #NW | $G_P$ (S) | $d_{nw}$ (nm) | $S_c$ (nm²) | $\sigma_P$ (S/cm) |
|---|---|---|---|---|
| #3 | 7.74x10$^{-12}$ | 50 | 39 | 9.93x10$^{-05}$ |
| #3 | 3.87x10$^{-13}$ | 50 | 39 | 4.96x10$^{-06}$ |
| #3 | 2.71x10$^{-12}$ | 50 | 39 | 3.47x10$^{-05}$ |
| #3 | 9.68x10$^{-12}$ | 50 | 39 | 1.24x10$^{-04}$ |
| #3 | 1.27x10$^{-9}$ | 50 | 39 | 1.63x10$^{-02}$ |
| #3 | 9.04x10$^{-11}$ | 50 | 39 | 1.16x10$^{-03}$ |
| #3 | 5.42x10$^{-12}$ | 50 | 39 | 6.95x10$^{-05}$ |
| #3 | 3.87x10$^{-12}$ | 50 | 39 | 4.96x10$^{-05}$ |
| #3 | 7.74x10$^{-13}$ | 50 | 39 | 9.93x10$^{-06}$ |
| #3 | 1.16x10$^{-11}$ | 50 | 39 | 1.49x10$^{-04}$ |
| #3 | 1.65x10$^{-11}$ | 50 | 39 | 2.11x10$^{-04}$ |
| #3 | 1.16x10$^{-12}$ | 50 | 39 | 1.49x10$^{-05}$ |
| #6 | 4.80x10$^{-10}$ | 60 | 39 | 7.39x10$^{-03}$ |
| #6 | 1.22x10$^{-09}$ | 60 | 39 | 1.88x10$^{-02}$ |
| #6 | 5.61x10$^{-12}$ | 60 | 39 | 8.64x10$^{-05}$ |
| #6 | 2.75x10$^{-11}$ | 60 | 39 | 4.23x10$^{-04}$ |
| #6 | 7.37x10$^{-11}$ | 60 | 39 | 1.13x10$^{-03}$ |
| #6 | 8.03x10$^{-11}$ | 60 | 39 | 1.24x10$^{-03}$ |
| #6 | 1.10x10$^{-11}$ | 60 | 39 | 1.70x10$^{-04}$ |
| #6 | 1.37x10$^{-11}$ | 60 | 39 | 2.11x10$^{-04}$ |
| #6 | 6.08x10$^{-10}$ | 60 | 39 | 9.36x10$^{-03}$ |
| #6 | 5.36x10$^{-10}$ | 60 | 39 | 8.25x10$^{-03}$ |
| #6 | 4.30x10$^{-11}$ | 60 | 39 | 6.61x10$^{-04}$ |
| #6 | 4.92x10$^{-11}$ | 60 | 39 | 7.56x10$^{-04}$ |

| #NW | $G_P$ (S) | $d_{nw}$ (nm) | $S_c$ (nm²) | $\sigma_P$ (S/cm) |
|---|---|---|---|---|
| #12 | 1.09x10⁻¹⁰ | 60 | 39 | 1.68x10⁻⁰³ |
| #12 | 3.55x10⁻¹⁰ | 60 | 39 | 5.46x10⁻⁰³ |
| #12 | 2.46x10⁻¹⁰ | 60 | 39 | 3.79x10⁻⁰³ |
| #12 | 1.92x10⁻¹⁰ | 60 | 39 | 2.95x10⁻⁰³ |
| #12 | 1.32x10⁻¹¹ | 60 | 39 | 2.02x10⁻⁰⁴ |
| #12 | 3.21x10⁻¹¹ | 60 | 39 | 4.94x10⁻⁰⁴ |
| #1 | 2.15x10⁻¹¹ | 60 | 39 | 3.31x10⁻⁰⁴ |
| #1 | 1.19x10⁻¹¹ | 60 | 39 | 1.83x10⁻⁰⁴ |
| #1 | 3.55x10⁻¹⁰ | 60 | 39 | 5.46x10⁻⁰³ |
| #1 | 5.87x10⁻¹² | 60 | 39 | 9.03x10⁻⁰⁵ |
| #1 | 3.24x10⁻¹⁰ | 60 | 39 | 4.98x10⁻⁰³ |
| #1 | 2.97x10⁻¹⁰ | 60 | 39 | 4.57x10⁻⁰³ |
| #1 | 5.57x10⁻¹¹ | 60 | 39 | 8.57x10⁻⁰⁴ |
| #1 | 6.61x10⁻¹¹ | 60 | 39 | 1.02x10⁻⁰³ |
| #1 | 1.77x10⁻¹¹ | 60 | 39 | 2.73x10⁻⁰⁴ |
| #1 | 7.81x10⁻¹¹ | 60 | 39 | 1.20x10⁻⁰³ |

**Table S7.** *Zero-bias perpendicular conductance $G_P$ from C-AFM data shown in Fig. 6a, NW diameter $d_{nw}$ from topographic AFM images, C-AFM tip contact area $S_c$ (see below) and calculated perpendicular conductivity $\sigma_p = G_P(S_c/d_{nw})$.*

## S11. C-AFM contact area.

The loading force was set at ~ 60 nN for all the I-V measurements, a lower value leading to too many contact instabilities during the I-V measurements. The contact radius, $r_c$, between the C-AFM tip and the NW surface, and the NW elastic deformation, δ, were estimated from a Hertzian model:[17]

$$r_c^2 = \left(\frac{3RF}{4E^*}\right)^{2/3} \tag{S1}$$

$$\delta = \left(\frac{9}{16R}\right)^{1/3}\left(\frac{F}{E^*}\right)^{2/3} \tag{S2}$$

with F the tip loading force (~60 nN), R the tip radius (25 nm) and E* the reduced effective Young modulus defined as:

$$E^* = \left(\frac{1}{E_{nw}^*} + \frac{1}{E_{tip}^*}\right)^{-1} = \left(\frac{1-v_{nw}^2}{E_{nw}} + \frac{1-v_{tip}^2}{E_{tip}}\right)^{-1} \tag{S3}$$

In this equation, $E_{nw/tip}$ and $v_{nw/tip}$ are the Young modulus and the Poisson ratio of the NWs and C-AFM tip, respectively. For the Pt/Ir (90%/10%) tip, we have $E_{tip}$ =

204 GPa and $\nu_{tip}$ = 0.37 using a rule of mixture with the known material data.[18] For the $W_{18}O_{49}$ nanostructures, we assumed a mean value of an effective Young modulus $E^*_{nw} = E_{nw}$ = 28 GPa from a series of mechanical measurements (3 points contact on suspended NWs using a contact mode AFM) on tungsten oxide nanowires with a diameter of ~ 100 nm (the Poisson ratio is not known).[19] With these parameters, we estimated $r_c$ ≈ 3.5 nm (contact area ≈ 39 nm²) and δ ≈ 0.5 nm.

## S12. Additional figure on electron transport mechanisms.

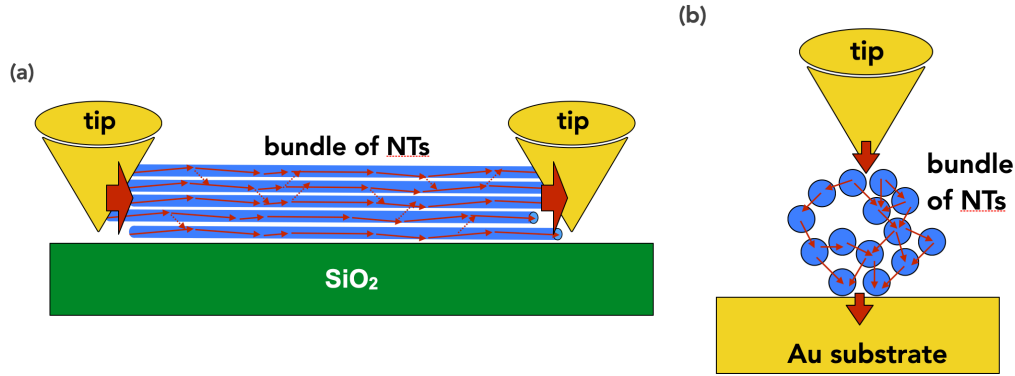

***Figure S10***. *(a) Scheme of the longitudinal electron transport in the bundle of NTs. The electron transport (for metal-like NTs) is likely due drift-diffusion along the individual NT (red solid arrows), with only a few tunnel hopping (dashed arrows) between adjacent NTs in the bundle. The transport along the NTs is limited by scattering events (at the NT surface and by defect/impurity in the NT). For simplicity, the NTs are schematized by solid blues cylinders) and only the two inner STM tips (see Fig. 4a in the main text) are visualized. (b) Scheme of the electron transport mechanism in the perpendicular direction. The electron transport across the NWs is mainly due to tunnel hopping (red solid arrows) between neighboring NTs. The topology of the ET pathways is more complex depending on how exactly a NT interacts with its neighboring NTs and how many they are around and how they are organized. For simplicity the cross-section of the NTs are schematized by blue circles.*